\documentclass[10,journal,compsoc]{IEEEtran}
\usepackage{microtype}                 
\PassOptionsToPackage{warn}{textcomp}  
\usepackage{textcomp}                  
\usepackage{mathptmx}                  
\usepackage{times}                     
\usepackage{tabu}                      
\usepackage{booktabs}                  
\usepackage{xcolor}

\usepackage{multirow}
\usepackage{amsmath}
\usepackage{amsthm}
\usepackage{hyperref}
\usepackage{float}
\usepackage{lineno,hyperref}
\usepackage{subeqnarray}
\usepackage{cases}
\usepackage{enumitem}
\usepackage{algorithm}
\usepackage{algorithmic}
\usepackage{array}
\usepackage{bm}
\usepackage{subcaption}
\usepackage{wrapfig}

\usepackage{graphbox}
\graphicspath{{../pdf/}{../jpeg/}}
\DeclareGraphicsExtensions{.pdf,.jpeg,.png, .jpg, .eps}

\ifCLASSOPTIONcompsoc
  \usepackage[nocompress]{cite}
\else
  \usepackage{cite}
\fi

\usepackage[normalem]{ulem}
\newcommand{\stkout}[1]{\ifmmode\text{\sout{\ensuremath{#1}}}\else\sout{#1}\fi}

\makeatletter
\newcommand*{\rom}[1]{\expandafter\@slowromancap\romannumeral #1@}
\makeatother

\begin{document}

\title{SGLDBench: A Benchmark Suite \\ for Stress-Guided Lightweight 3D Designs}

\author{Junpeng~Wang, Dennis~R.~Bukenberger, Simon~Niedermayr, Christoph~Neuhauser, Jun~Wu, and~R\"udiger~Westermann
\IEEEcompsocitemizethanks{\IEEEcompsocthanksitem 
J. Wang, D. Bukenberger, S. Niedermayr, C. Neuhauser and R. Westermann are with the Computer Graphics \& Visualization Group, Technische Universit{\"a}t M{\"u}nchen, Garching, Germany.\protect\\
E-mail: \{junpeng.wang, dennis.bukenberger, simon.niedermayr, christoph.neuhauser, westermann\}@tum.de.} 
\IEEEcompsocitemizethanks{\IEEEcompsocthanksitem 
J. Wu is with the Department of Sustainable Design Engineering, Delft University of Technology, Delft, The Netherlands. \protect\\
E-mail: j.wu-1@tudelft.nl.} 
}

\IEEEtitleabstractindextext{%
\begin{abstract} We introduce the Stress-Guided Lightweight Design Benchmark (SGLDBench), a comprehensive benchmark suite for applying and evaluating material layout strategies to generate stiff, lightweight designs in 3D domains. SGLDBench provides a seamlessly integrated simulation and analysis framework, including six reference strategies and a scalable multigrid elasticity solver to efficiently execute these strategies and validate the stiffness of their results. This facilitates the systematic analysis and comparison of design strategies based on the mechanical properties they achieve. SGLDBench enables the evaluation of diverse load conditions and, through the tight integration of the solver, supports high-resolution designs and stiffness analysis. Additionally, SGLDBench emphasizes visual analysis to explore the relationship between the geometric structure of a design and the distribution of stresses, offering insights into the specific properties and behaviors of different design strategies. SGLDBench's specific features are highlighted through several experiments, comparing the results of reference strategies with respect to geometric and mechanical properties.

\end{abstract}

\begin{IEEEkeywords}
Topology optimization, lattice infill, lightweight design, simulation design.
\end{IEEEkeywords}}

\maketitle

\IEEEdisplaynontitleabstractindextext

\IEEEpeerreviewmaketitle


\section{Introduction}
Topology optimization (TO) and functionally graded lattice infill are primary strategies for designing mechanically sound, lightweight structures, i.e., structures with high stiffness (corresponding to a low compliance, or degree of deformability) under applied loads. TO determines the optimal material distribution within a given design domain to achieve a desired structural performance, such as maximizing stiffness, while satisfying constraints like material use~\cite{Bendsoe1989SO, Allaire2004JCP}. Functionally graded lattice infill refers to a design approach in which the density, size, shape, or material properties of the lattice structure vary spatially within a 3D object to meet specific performance requirements~\cite{Wu2021SMO}. For beam-based lattices, the final design can be represented by a graph or grid composed of polyhedral cells, each constructed from individual edges.

TO, in its basic form, does not consider the geometric structure of the resulting material layout but aims to achieve the highest possible stiffness. Lattice infill design strategies, in principle, share this goal, by tailoring the lattice layout based on the stress distribution. 
Stress is a measure of the internal forces that develop within a material when it is subjected to external loads and quantifies the intensity of these forces at a specific point in the material. 
The structural rigidity of an infill increases when the material aligns with the orthogonal principal stress directions of the object under load~\cite{pedersen1989optimal}. These directions, corresponding to the eigenvectors of the 3x3 stress tensor, indicate the normal stresses acting on specific planes within where shear stresses are zero. At the limit of material volume, considering these directions for the lattice layout results in microstructures resembling quads or hexahedra~\cite{michell1904lviii, Wu2021TVCG}.

Each lattice infill design strategy, however, involves additional considerations that may compromise stiffness. These include achieving geometric properties such as regularity (i.e., variation in element type), uniformity (i.e., variation in element size), or space-fillingness to enhance robustness, as well as purely aesthetic features~\cite{plocher2019review, Wu2021SMO, gil2020reinforcement, bacciaglia2019systematic}. 

Moreover, generating 3D domain-filling lattice structures that align with major stress directions presents significant challenges. This difficulty arises from the existence of degenerate points~\cite{delmarcelle1994topology} (or degenerate regions in 3D domains~\cite{wang2020globally, hergl2021visualization, hung2023global}) where the stress tensor has repeating eigenvalues, making the principal stress directions indeterminate. As a result, integrability conditions are violated, and consistent domain parameterizations cannot be computed~\cite{Stutz2020SMO}.

To assist users in selecting the right 3D lightweight design strategy for various use cases, and to help researchers identify open research questions, a benchmark suite for generating, analyzing, and comparing the results of different strategies is essential.

\begin{figure*}[t]
   \centering
   \includegraphics[width=0.98\linewidth,trim=0.0cm 0.0cm 0.0cm 0.0cm, clip=true]{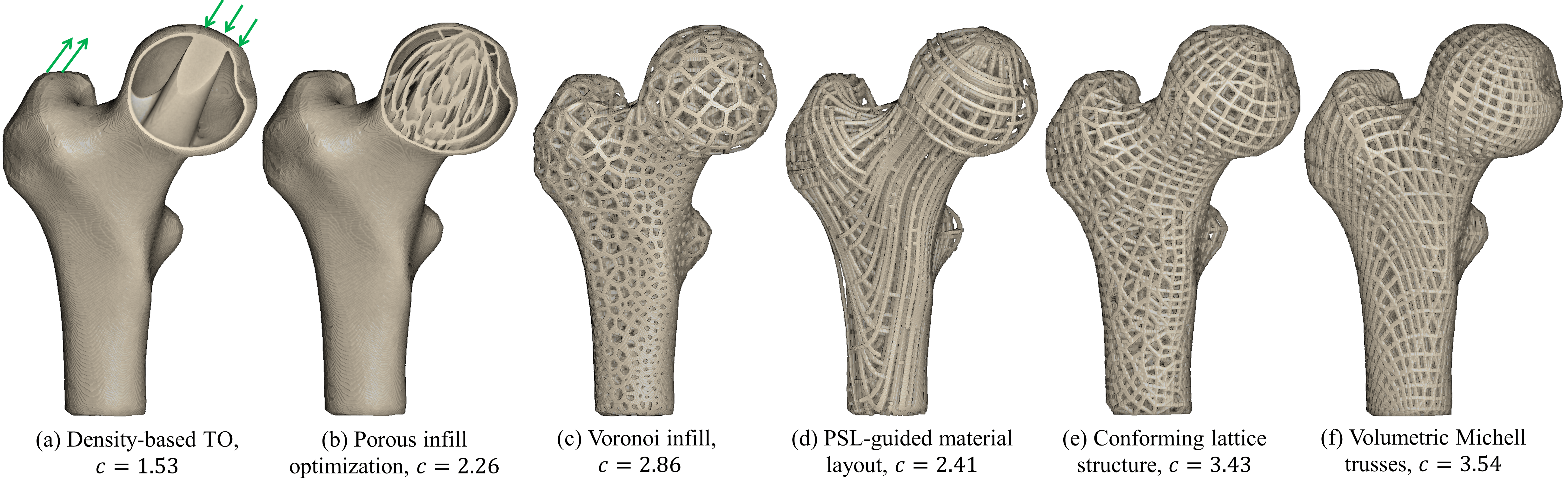} 
   \caption{Infill designs computed using the strategies provided by SGLDBench.
   All designs consume roughly the same amount of material and are subject to the same boundary conditions: the bottom of the design domain is fixed, and the loads are indicated by green arrows in (a). Below each design, its compliance $c$ is provided. For visualization, isosurface volume rendering is used. All designs are coated with fully solid boundary elements of consistent thickness. In (c) to (f), these elements are peeled away to reveal the infills.}
   \label{fig:teaser}
 \end{figure*}
 
A few benchmark papers have addressed issues such as special solvers for TO~\cite{rojas2015benchmarking}, benchmarks for 2D TO in specific load cases~\cite{valdez2017topology}, practices that should be considered when performing TO~\cite{sigmund2022benchmarking}, as well as the mechanical soundness of simple lattice infills such as orthogonal grids and shells~\cite{PERNET2022682}. Different architectures for multidisciplinary design optimization have been reviewed and compared~\cite{martins2013multidisciplinary}, and the design and structural optimization of lightweight design has been discussed, especially in the context of additive manufacturing~\cite{plocher2019review}. A review of uniform and non-uniform lattice structures such as foams and honeycombs sheds light on their properties and methods for designing and optimizing such structures~\cite{tamburrino2018design,pan2020design}. Unit cell lattices comprising structures made of a single type of cells have been researched~\cite{seharing2020review}, and the properties of certain types of lattice infills regarding 3D printing processes have been discussed~\cite{maconachie2019slm}. The combination of TO and micro element-based lattice infills have resulted in structures exhibiting anisotropic mechanical properties~\cite{zhang2021novel, zhu2017two}. There is no benchmark that allows researchers and users to efficiently compute 3D designs with different strategies and to effectively compare the results with respect to their mechanical and structural properties. 

We introduce \emph{SGLDBench} to address this gap. It provides a comprehensive investigation of the combination of boundary shapes and conditions with lightweight design strategies in 3D domains. 
This is facilitated by a seamlessly integrated, MATLAB-based simulation and analysis framework offering the following key features:

\begin{itemize} 
\item \textbf{Selection of Lightweight 3D Design Strategies}: The benchmark includes various strategies, enabling comparisons between TO and lattice infill, as well as studies of new scenarios and designs. For a 3D human femur under load, Fig.~\ref{fig:teaser} shows visualizations of the infill designs computed by SGLDBench.
\item \textbf{Material Layout Generation}: A central feature of the benchmark is the voxelization of complex infills into a Cartesian simulation grid. This ensures consistent comparisons of lightweight designs---whether represented as a material field, mesh, or edge graph---with respect to their mechanical properties. 
\item \textbf{Simulation Suite}: SGLDBench provides a MATLAB-interfaced simulation framework with an efficient multigrid solver for generating high-resolution stress fields and assessing the stiffness of a design efficiently. 
\item \textbf{Visual Design Analysis}: A fast volume visualization module accompanies the benchmark, offering visual feedback on a design’s shape, stress distribution, and material alignment before and after layout optimization. \end{itemize}

We chose MATLAB as the working environment due to its widespread use in computational design. SGLDBench leverages MATLAB's simulation and visualization capabilities for design generation and analysis. New design strategies can be integrated either through MATLAB programs or by using MATLAB’s functionality to call executables or Python programs from other codebases via inline calls. All SGLDBench-specific operations have been implemented in MATLAB or rely on publicly available MATLAB, C++, or Python programs. SGLDBench also uses external libraries for core operations such as meshing and voxelization. The visualization module is implemented in WebGL, allowing it to function either as a standalone viewer in a web browser or via inline calls to MATLAB’s viewing functionality. Upon acceptance, the entire codebase for SGLDBench will be made publicly available.

\begin{figure*}[t]
   \centering
   \includegraphics[width=0.98\linewidth,trim=0.0cm 0.0cm 0.0cm 0.0cm, clip=true]{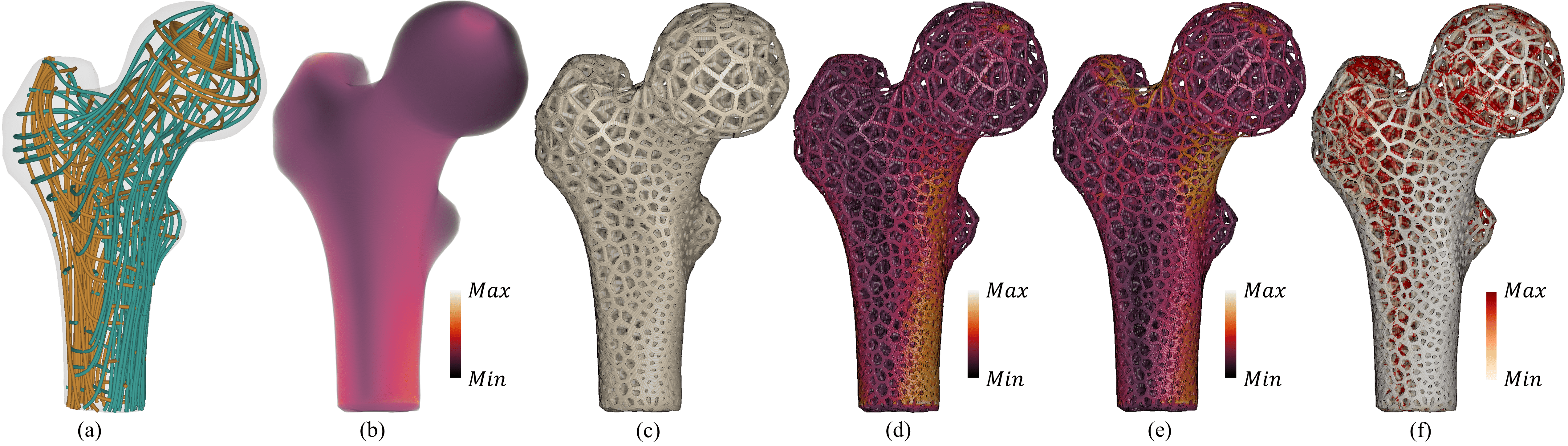} 
   \caption{
   SGLDBench's visual analysis options. (a) Major (brown) and minor (green) PSLs according to boundary conditions from Fig.~\ref{fig:teaser}. (b) Direct volume rendering of scalar von Mises stresses in the solid domain. (c) Voronoi infill (c = 2.86). (d) Infill with color-coded von Mises stresses. (e) Same as (d), but different forces apply and von Mises stresses change (c = 4.14). (f) Comparative visual stress analysis showing the misalignment between the major stress directions in the solid and the infill under the same loads.  
   }
   \label{fig:teaserVis}
 \end{figure*}

\section{SGLDBench's Functional Structure}
We begin by introducing the key functionality of SGLDBench. SGLDBench provides an interface to enable users to specify the \textbf{boundary conditions}, which include the boundary of a domain, the applied loads, and the fixed boundary regions.
The domain is then voxelized, meaning it is discretized into a Cartesian grid, with per-voxel properties assigned based on the boundary conditions and material properties. We refer to this configuration as a \textbf{preset}.

Using the selected preset, SGLDBench \textbf{simulates the object's internal stress field} via a multigrid finite-element elasticity solver~\cite{dick2011real,wu2015system,liu2018narrow,aage2017giga}. While some design strategies require repeated solver executions to iteratively optimize the material layout, others generate an infill structure guided by the principal stresses in the initial field.

Users can choose from \textbf{six layout strategies} for computing an \textbf{infill design}: density-based TO ~\cite{Sigmund2001SMO}, porous infill optimization~\cite{Wu2018TVCG}, Voronoi infill~\cite{lu2014build}, stress-line-guided material layout~\cite{WWW2022stressTrajectory}, conforming lattice structures~\cite{gao2017robust}, and volumetric Michell trusses~\cite{arora2019volumetric}.
 
Our selection has not been made with the intention to favor any of these methods, but to reveal the specific characteristics of 3D design strategies following different objectives. The methods span the spectrum from purely stiffness-based optimization to geometry-aware infill generation. We select \textbf{density-based TO} as a representative of various TO approaches. It serves as a reference for the stiffness that can be achieved. \textbf{Porous infill optimization}, while not explicitly using the principal stress directions, results in wall-like structures that largely agree with two of these directions. \textbf{Voronoi infill} considers only the stress magnitude but not its principal directions. In contrast, material layouts guided by the \textbf{Principal Stress Lines (PSLs)} follow exactly the mutually orthogonal principal stress trajectories in the initial solid domain. PSL-guided infills serve as a reference conveying these directions, even though the final designs are not connected in general. \textbf{Conforming lattice structures} and \textbf{volumetric Michell trusses} aim at finding a balance between stress alignment and geometric regularity of the designs. While the first approach favors stress alignment and, therefore, needs to resort to an edge-graph structure, the latter approach strives for a pure hexahedral mesh and, therefore, needs to sacrifice stress alignment.

While TO and porous infill generate a material field, other methods compute a lattice structure composed of edges and nodes. SGLDBench \textbf{voxelizes} these structures into a material field on a Cartesian grid with the same resolution as the initial preset. Using the material field and boundary conditions, the elasticity solver computes the \textbf{compliance of the design}. When using iterative optimization methods, the compliance history is recorded and can be visualized.

SGLDBench supports different \textbf{visualization options} to inspect a 3D design. The principal stress directions in a stress field are visualized using \textbf{PSL-guided trajectory visualization} implemented in MATLAB programs~\cite{wang20223d}. These programs are accessible through SGLDBench's interface and allow users to customize the number and appearance of the visualized trajectories (see Fig.~\ref{fig:teaserVis}a).

From the principal stresses the \textbf{scalar von Mises stresses} are computed. The von Mises stress is commonly used in engineering and materials science to predict yielding in ductile materials under load. It provides a single value that reflects the combined effect of all stress components acting on a material. This scalar field can then be visualized with SGLDBench's \textbf{WebGL-based visualization module} for enhanced rendering performance (Fig.~\ref{fig:teaserVis}b). An infill structure is rendered as an iso-surface in the material field (Fig.~\ref{fig:teaserVis}c), and it can be color-coded with the von Mises stress to reveal local stress concentrations (Fig.~\ref{fig:teaserVis}d).

Additionally, a \textbf{variable load structural analysis} can be performed, by loading a design with forces different from those for which it was initially optimized. This functionality enables users to evaluate the robustness of a design under different loading conditions. The optimized design can be color-coded with the von Mises stresses occurring under the new load conditions (Fig.~\ref{fig:teaserVis}e).

Furthermore, SGLDBench offers tailored visualization options to examine how the mechanical properties of the initial solid and the generated infill design have changed. Direct volume rendering is used with a predefined color transfer function to visualize the \textbf{per-voxel stress deviations} in the final design relative to those in the initial solid body (see Fig.~\ref{fig:teaserVis}f).

\section{Components of SGLDBench}
We describe here the most important features and operations of SGLDBench, including descriptions of the supported TO and lattice infill methods. The use of SGLDBench is demonstrated in the accompanying videos.

\subsection{Domain specification}
SGLDBench simulates a stress field in the design domain using Finite Element Analysis (FEA). This requires discretizing the domain into finite elements and specifying boundary conditions. SGLDBench uses a hexahedral finite-element representation to facilitate the use of scalable geometric multigrid solvers. Therefore, SGLDBench first converts an initial object representation to a hexahedral simulation grid.

\noindent \textbf{\emph{Voxelization.}}
The simulation grid is created by voxelizing the simulation domain. The user provides the domain boundary as a closed triangular mesh. SGLDBench utilizes MATLAB's voxelization capabilities~\cite{Adam2024} with a user-defined voxel resolution to compute a solid voxelization. For complex-shaped simulation domains, the voxels are classified as solid or void, depending on the centroid of the voxel. Alternatively, users can upload a 3D voxel grid that discretizes the domain and marks each voxel as solid or void. Void elements are excluded from the finite element analysis.

Voxels with at least one of their 26 neighboring voxels classified as void are designated as boundary voxels. SGLDBench applies a dilation operation to expand the boundary by assigning any voxel adjacent to an initial boundary voxel as a new boundary voxel. This voxelized object serves as the foundation for all subsequent operations in SGLDBench.

\noindent \textbf{\emph{Boundary Conditions. }} 
The boundary conditions define where the object is fixed and how loads are applied. Fixations and forces are assigned to the nodes of the boundary elements. The user specifies the extent and position of an axis-aligned box or sphere, and SGLDBench automatically fixes or applies the specified loads to all boundary nodes within this region. Similarly, the nodes to be reset can be selected in the same manner. When all nodes of a finite element are fixed, the element becomes rigid and does not respond to any loads.

\begin{figure}[t]
    \centering
    \includegraphics[width=0.98\linewidth, trim=0.0cm 0.0cm 0.0cm 0.0cm, clip=true]{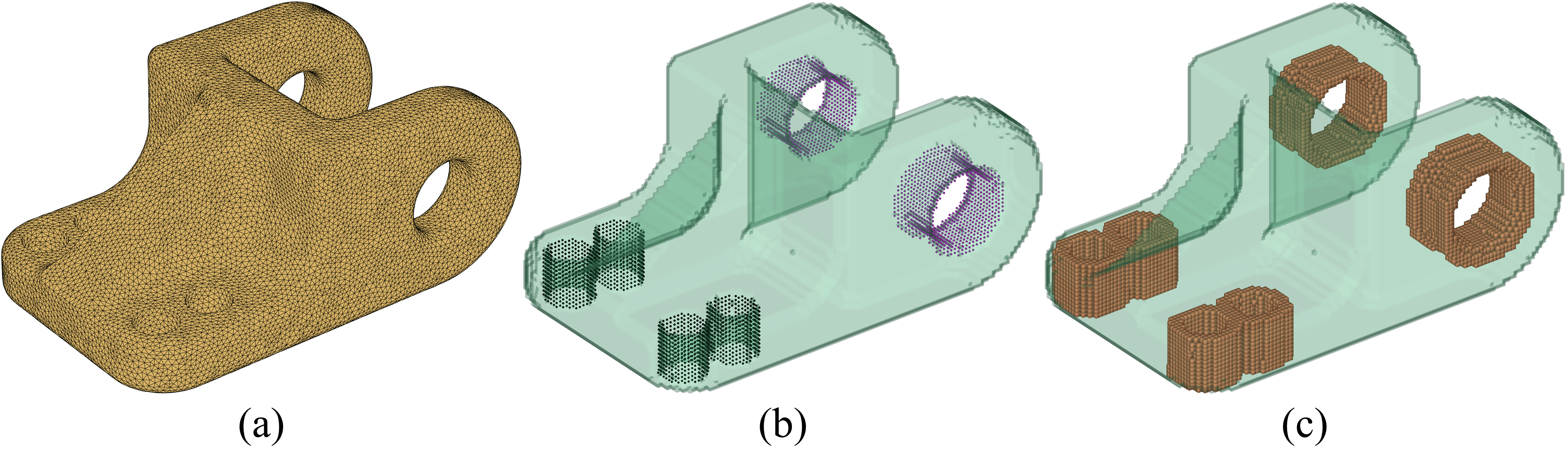}
    \caption{
    (a) A triangle mesh defines the domain boundary. (b) After voxelizing the domain, the user specifies boundary forces and fixes grid vertices. Black and violet dots indicate fixed vertices and vertices subjected to an external force, respectively. (c) Passive elements are shown in brown.
    }
    \label{fig:Voxelization} 
\end{figure}

\noindent \textbf{\emph{Passive Elements. }} 
Passive elements are used to preserve specific geometric features, such as object boundaries or notches for mounting connections. Passive elements remain solid throughout the optimization process and contribute to the stiffness of the structure. SGLDBench supports two general methods for specifying passive elements:
Setting all boundary elements as passive, which is common in infill design problems, and
setting elements with loaded or fixed nodes as passive to preserve geometric features during optimization.
The dilation operation can also be used to enlarge passive structures as needed. Figure~\ref{fig:Voxelization} illustrates the transformation from a boundary mesh to a voxel model, including boundary conditions and passive elements.

\subsection{Stress simulation} \label{subsec:stressSim}

At the core of TO and lattice infill methods is the numerical simulation of a stress tensor field using the selected boundary conditions and material properties. SGLDBench provides a MATLAB-interfaced C++ implementation of a multigrid elasticity solver to efficiently simulate the stress field.

The implementation employs a geometric multigrid solver as a preconditioner for a conjugate gradient method to solve a sparse linear system of equations, i.e., $KU = F$. The global stiffness matrix $K$ is assembled from the element stiffness matrices under the assumption of a linear material law. The computation of the element stiffness matrices accounts for the stiffness tensor, which reflects material properties, and the strain matrix, which expresses the strain-stress relationship. $U$ represents the static displacement vector in response to the external loads $F$. Several prior works have addressed the efficient assembly of the system matrix $K$ and the specific adaptations of numerical solvers for linear elasticity simulations in TO  ~\cite{dick2011real,wu2015system,liu2018narrow,aage2017giga, traff2023simple}. 

SGLDBench's multigrid implementation is primarily based on the work of Wu et al.~\cite{wu2015system}, utilizing on-the-fly numerical stencil assembly and multigrid interpolation and restriction across multiple levels simultaneously. However, the implementation has been adapted for MATLAB running on a CPU, resulting in changes to the internal data and computation layouts.

\begin{figure}[t]
    \centering
    \includegraphics[width=0.98\linewidth, trim=0.0cm 0.0cm 0.0cm 0.0cm, clip=true]{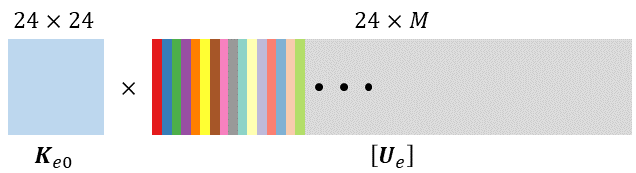}
    \caption{Element-based computation layout. $M$ is the number of hexahedral simulation elements, $K_{e0}$ is the generic element stiffness matrix and $U_e$ is the element-wise displacement matrix. Different colors represent the 8 node-based displacement vectors per element. With each column $l_i$ in $U_e$, MATLAB computes the product $K_{e0} \cdot l_i$}. 
    \label{fig:matrixFree} 
\end{figure}

Firstly, SGLDBench uses MATLAB's built-in Cholesky solver rather than the TAUCS library’s Cholesky solver for solving the linear system on the coarsest multigrid level. Secondly, it transitions from a matrix-free node-based computation layout to a matrix-free element-based layout to take advantage of MATLAB's efficient matrix-vector operations. Instead of assembling stencils per grid vertex on-the-fly using indexed memory access operations, SGLDBench constructs a generic element matrix and utilizes MATLAB to compute the products of this matrix with the 8-node displacement vectors of each element. Since the stiffness matrix of an element with density $\rho$ is obtained by correspondingly scaling the generic stiffness matrix, the final results only need to be scaled accordingly. 
For each element, the 8 displacement vectors at the vertices are organized into columns of an element-wise displacement matrix, as illustrated in Fig.~\ref{fig:matrixFree}. MATLAB then computes all matrix-vector products between the generic element matrix and each column in the displacement matrix efficiently.

\begin{figure*}[t]
    \centering
    \begin{subfigure}[b]{0.48\textwidth}
    \includegraphics[width=\linewidth]{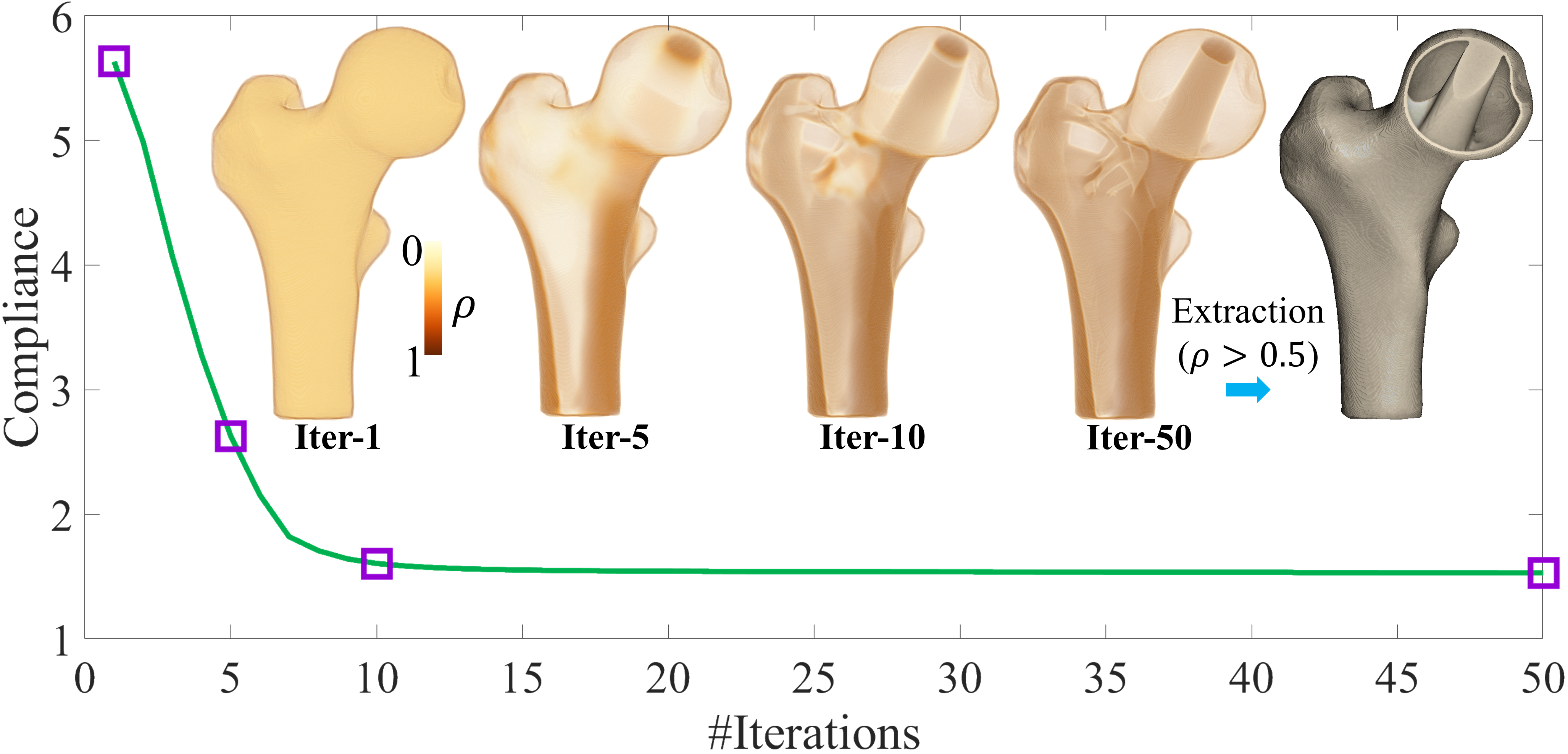}
    \caption{}
    \end{subfigure}
    \begin{subfigure}[b]{0.48\textwidth}
    \includegraphics[width=\linewidth]{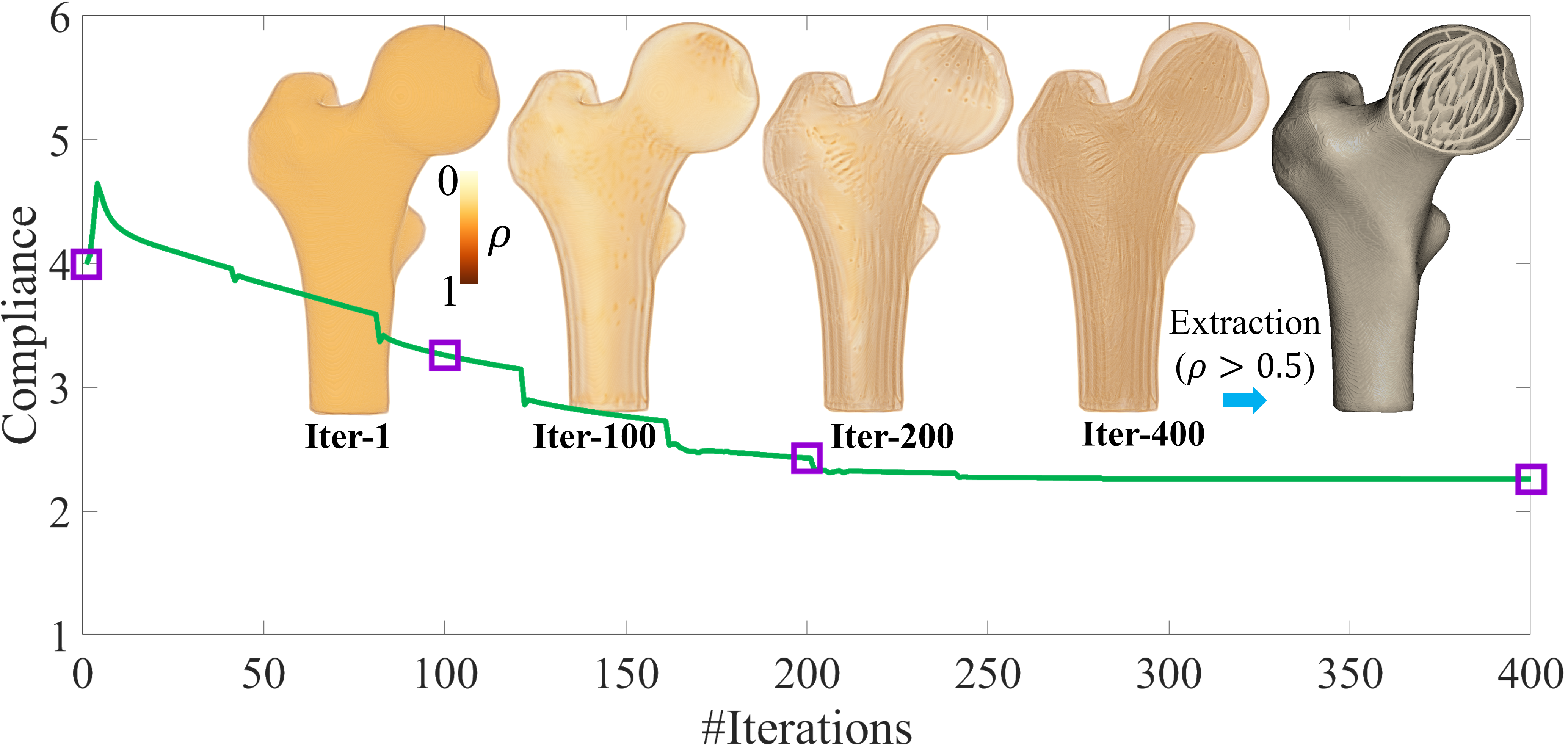}
    \caption{}
    \end{subfigure}
    \caption{
    Material layout optimization with density-based TO (a) and porous infill optimization (b). Compliance history is depicted by the green curves, with violet squares indicating shown states.
    }
    \label{fig:demo_topopti_GL} 
\end{figure*}

\subsection{Infill computation} 
For each of the six lightweight design methods provided by SGLDBench, the user selects specific parameters and lets SGLDBench compute the material layout. SGLDBench provides MATLAB code for density-based TO, porous infill optimization, PSL-guided material layout and volumetric Michell trusses. To generate a Voronoi infill, MATLAB calls a Python script including precompiled libraries. Conforming lattice structures are generated by compiling code from a publically available repository and running the executable from MATLAB with the required parameters. 

\subsubsection{Topology optimization}
TO minimizes the compliance of a material layout under the constraint of applied forces and a prescribed global material consumption. SGLDBench implements the density-based TO approach ~\cite{Sigmund2001SMO, liu2014efficient}. To formulate the minimization problem over a discrete set of elements $e$ with densities $\rho_e$, a hexahedral finite element discretization of a linear elastic solid material is generated from the voxelized geometry. The object’s compliance $c$ is computed by summing the strain energy over all material elements, i.e., 
\begin{equation} \label{eqn:compliance}
    c = U^{\text{T}}KU.
\end{equation}
The lower the compliance, the higher the object's stiffness.

With selected measure of a material's ability to deform under an applied stress, i.e., the Young’s modulus $E_0$ of the solid ($\rho_e = 1$), and the linear material law, TO proceeds in three steps: 1) A large linear system is solved using the MATLAB multigrid implementation to compute the force-induced displacements of the hexahedral vertices. 2) The derivatives of the total strain energy $c$ and the total volume $V$ with respect to the elements' densities $\rho_e$ are computed and used to guide the material distribution to maximize stiffness. 3) The design is updated according to the computed sensitivities. These steps are repeated until the change in material distribution is below a threshold or the number of iterations reaches the prescribed maximum iterations. The computational pipeline for density-based TO is mainly written in MATLAB, using our optimized C++ code for solving the linear system and updating the design variables.

Density-based TO takes the available material budget $V_0$ as the constraint, known as the global volume constraint. Thus, the problem-specific constraint function for a material layout $\phi$ using $n$ hexahedral elements is  
\begin{equation} \label{eqn:GlobalVF}
    g \left(\phi \right) = \sum \rho_e - nV_0 \leq 0
\end{equation}
We use the so-called Solid Isotropic Material with Penalization (SIMP) model, where a non-zero constant minimum value represents the background stiffness of the void region. In each optimization iteration, the design variables are updated within a prescribed step size through a gradient-based optimizer using the optimality criteria method ~\cite{Sigmund2001SMO}. After optimization, the design variables shall converge to a (near-)binary layout that indicates the spatial material distribution. 

It's worth mentioning that several auxiliary processes are also introduced in practical TO for good design quality. For instance, the density-based filtering to counteract numerical instabilities and the Heaviside projection to promote the generation of a binary design, where the proxy density value of each voxel is encoded by the projected value of the filtered value of the design variable~\cite{wang2011projection}.

Fig.~\ref{fig:demo_topopti_GL}a shows the optimized shapes after different optimization iterations. The optimization produces a mono-scale design comprising mainly a thick resistant strand along the maximum stress directions. 

\subsubsection{Porous infill optimization}
Porous infill optimization is an extension of density-based TO that generates porous substructures distributed across the design domain. This is achieved by replacing the global volume constraint with local volume constraints, which prevent material from accumulating and forming dense, solid regions.

The global volume constraint restricts the total material consumption within the entire simulation domain. The local volume constraint imposes an upper bound $V_e$ on the percentage of solid voxels within a prescribed neighborhood of voxel $e$. These local constraints ensure a more evenly distributed material layout, promoting porosity and lightweight design. Beyond this adjustment, the optimization process largely follows the approach used in density-based TO.

For the material around each voxel $e$, the local volume constraint leads to the constraint function  
\begin{equation} \label{eqn:LocalVF}
    g \left(\phi_e \right) = \frac{\sum_{i\in N_e}\rho_i}{\sum_{i\in N_e}1} - V_e \leq 0. 
\end{equation}
$N_e$ defines the voxel neighborhood that is considered, i.e., 
\begin{equation} \label{eqn:LocalVFRange}
    \quad N_e = \{i| \parallel x_i-x_e \parallel_2 \leq R_e \}, \:\: \forall e.
\end{equation}
$R_e$ is the radius of a spherical region centered at a voxel's center, It defines the area within which local material accumulation is measured. SGLDBench's implementation of density-based TO in MATLAB has been extended to include the specific constraint function for porous infill optimization. The optimization process uses the Method of Moving Asymptotes (MMA) ~\cite{svanberg1987method} as the optimizer to iteratively update the design variables.

Fig.~\ref{fig:demo_topopti_GL}b illustrates the optimization process of porous infill optimization. Unlike standard TO, porous infill optimization generates a space-filling, multi-scale design. These designs generally exhibit lower stiffness compared to those produced by TO with a global volume constraint, as some material is deposited in regions that do not significantly contribute to overall stiffness.
However, such designs are typically more robust under varying load conditions and localized damage~\cite{Wu2018TVCG, WWW2022stressTrajectory}. Additionally, porous infill optimization in 3D tends to form wall-like structures aligned with the major and minor principal stress directions, as shown in Fig.~\ref{fig:teaser}b.

\subsubsection{Voronoi Infill}
3D Voronoi infills are generated by computing an initial Delaunay tetrahedralization, based on a set of samples ($S$) following a stress-based distribution density.
Therefore, more samples are generated in regions of higher stress, whereas the sampling density is lower in less stress-critical regions~\cite{lu2014build}. The Voronoi mesh follows as the dual of the Delaunay complex. Procedural infill optimization techniques building upon similar concepts have been proposed for additive manufacturing~\cite{martinez2016procedural,k-nearest-foams}.

\noindent \textbf{\emph{Graded Sampling. }}
        In SGLDBench, $S$ is initialized with a small set of auxiliary samples, equally distributed on a sphere, fully enclosing the input object.
        Further initial samples are added from the set of vertices of the input object's hull.
        Then, the input tetrahedral mesh is used as sampling domain, where $S$ is iteratively updated in a progressive Poisson disk sampling scheme until no further samples can be added.
        For improved performance, this is realized using batches of $n$ samples per iteration and organizing $S$ in a kd-tree.
        Radii for Poisson disks are interpolated at their sample positions based on the von Mises stress field $\sigma_\text{vm}$, using the mapping
        \begin{equation} \label{eqn:vmStressMapping}
            R = \text{icdf}(\sigma_\text{vm}) \cdot (\hat{r} \rho - \hat{r}) + \hat{r}
        \end{equation}
        where $\hat{r}$ the size of the largest radius (determined as a fraction of the objects bounding box diagonal length) and $\rho \in (0,1]$ gives the ratio of smallest to largest radii.
        As the von Mises stress has an arbitrary range from smallest to largest values with spatially varying concentrated extremes, we normalize and homogenize the field using an \textit{inverse cumulative distribution function} (icdf).

\noindent \textbf{\emph{Restricted Delaunay / Voronoi.}}
        The Delaunay complex or Voronoi diagram resulting from the generated samples are not natively limited to the design domain, i.e., the object's outer boundary.     
        As $S$ includes vertices from the object's hull as samples, the Delaunay complex is restricted to the object's shape by simply excluding Delaunay simplices outside of the object using robust winding numbers~\cite{jacobson2013robust}.
        Due to the dual nature of the Voronoi diagram and the Delaunay complex, the Voronoi cells of such hull vertex samples always transcend the object's outer boundary.
        Therefore, Voronoi cells crossing the outer hull are cut and clipped~\cite{sutherland1974reentrant, edelsbrunner1994triangulating} such that only their inner part remains, cells fully outside are omitted.

The edges of a Voronoi infill do not follow the stress directions in the initial solid object.
Whereas the Delaunay criterion guarantees the most regular simplices when applied on the available sampling points, there is no trivial control for edge directions in the Voronoi graph, for instance, to construct Voronoi infills with controlled elasticity ~\cite{martinez2016procedural} or constraint alignment of the Voronoi edges ~\cite{martinez2018polyhedral}. 

\begin{figure}[t]
    \centering
    \includegraphics[width=0.98\linewidth, trim=0.0cm 0.0cm 0.0cm 0.0cm, clip=true]{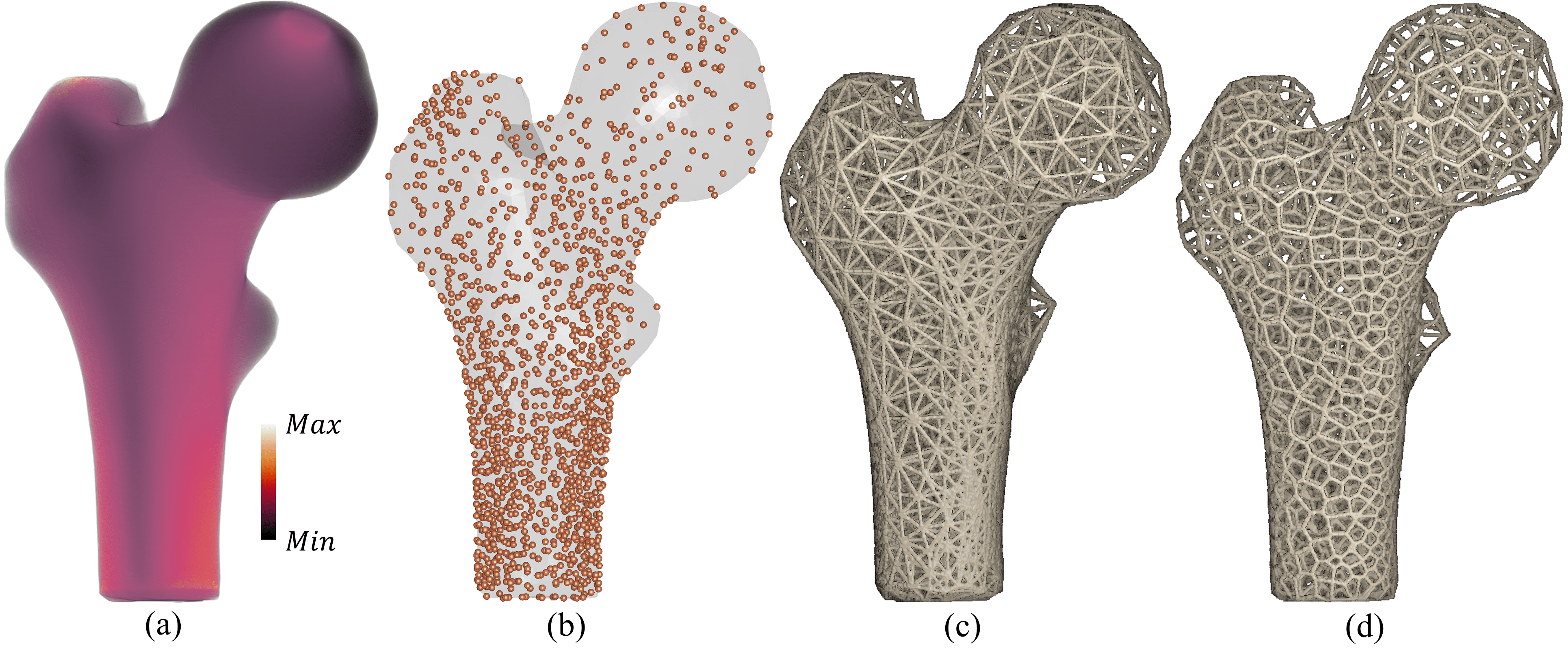}
    \caption{Stress-aware Voronoi infill. (a) Scalar von Mises stress field. (b) Point cloud with stress-aware density. (c) Tetrahedral Delaunay mesh from (b). (d) Voronoi infill from (c).}
    \label{fig:demo_Voronoi} 
\end{figure}

SGLDBench provides Voronoi infill generation via Python due to the easy accessibility of required functionality.
The SciPy~\cite{virtanen2020scipy} library is used to generate the Voronoi and Delaunay graphs using Qhull~\cite{barber1996quickhull} and further provides the kd-tree acceleration structure for the Poisson disk sampling. Our code includes fallback methods required for restricting the graph structure to the object domain. The pipeline is illustrated in Fig.~\ref{fig:demo_Voronoi}.

\subsubsection{PSL-guided lattice infill}

In the seminal work by Michell ~\cite{michell1904lviii} on stiffness-optimal lightweight design, it was conjectured that a stiffness-optimal structure should bear only normal stresses. This means that the sub-structures of such a design align with the principal stress directions. This is known as \emph{Michell's Theory}, which has been considered since then in various lightweight design methods. 

The most straightforward approach to create an infill that considers the principal stress directions in the loaded solid domain is to deposit material along the PSLs.
When using line seeding strategies to obtain an as uniform as possible and domain filling distribution of PSLs ~\cite{kwok2016structural, WWW2022stressTrajectory}, PSL-guided infills show surprisingly good mechanical properties in 2D domains~\cite{wang2023trajectory}. In 3D domains, however, many PSLs do not significantly contribute to the infill's stiffness, and PSLs might travel through space over a long distance before they intersect with any other PSL or attach to the boundary (see Fig.~\ref{fig:demo_ConformingLat}a).  

On the other hand, when depositing material by voxelizing lines with a selected thickness, the thicker the PSLs are, the more connections are generated. This increases significantly the mechanical performance, and it often results in infills that show superior mechanical properties. In addition, since the stress field needs to be computed only once in the solid domain, the computational complexity is significantly reduced.

SGLDBench uses the publically available MATLAB backend of 3D-TSV~\cite{wang20223d} to generate PSLs in a 3D stress tensor field. It generates a domain-filling and evenly-spaced set of PSLs. The method starts from a set of domain-filling seed points and iteratively creates PSL from these seeds. All remaining seeds in a certain distance to the PSL are removed to control the sparseness of the resulting PSL distribution.
The thickness of PSLs is selected by the user, and for a selected thickness the density of seeded PSLs is iteratively increased until the given volume budget is roughly reached. 
A PSL might enter into a region where the three principal stress directions are not uniquely defined and exchange their orientation, i.e., around so-called degenerate points~\cite{delmarcelle1994topology}. Tracing a PSL stops if the resulting directional change exceeds a given limit, and the PSL is removed to avoid wasting material.

\subsubsection{Conforming lattice structure}

Conforming lattice structure originates from the geometry-based structural dehomogenization presented by Wu~\emph{et al}~\cite{Wu2021TVCG}. Structural dehomogenization\cite{Pantz2008JCO, Groen2018IJNME,Allaire2018CMA} diverges from the fine-resolution simulation and optimization used by density-based TO and porous infill optimization by adopting a multi-scale strategy:\\
    \noindent \textbf{Homogenization-based TO.} During the initial optimization, a coarse-scale representation is tuned to approach the optimal distribution of material across a structure.
    This structure doesn't represent the exact material layout but rather provides a set of specifications to guide the optimal material layout.
    
    \noindent \textbf{Dehomogenization.} Once the optimized specifications are found, dehomogenization is the process of converting coarse, homogenized results into detailed, fine-scaled structures.
    This involves creating actual geometric elements (trusses, lattices, microstructures) realizing the material properties and orientations suggested by the homogenized result. 
    
Homogenization-based TO provides an orthotropic direction field as the optimized specifications. Dehomogenization extracts a conforming lattice structure with edges aligning with the direction field, and aspect ratios or sizes conveying the associated properties of the corresponding directions. The optimal directions are given by the principal stress directions of the homogenization-based structure layout. 

\begin{figure}[t]
    \centering
    \includegraphics[width=0.98\linewidth, trim=0.0cm 0.0cm 0.0cm 0.0cm, clip=true]{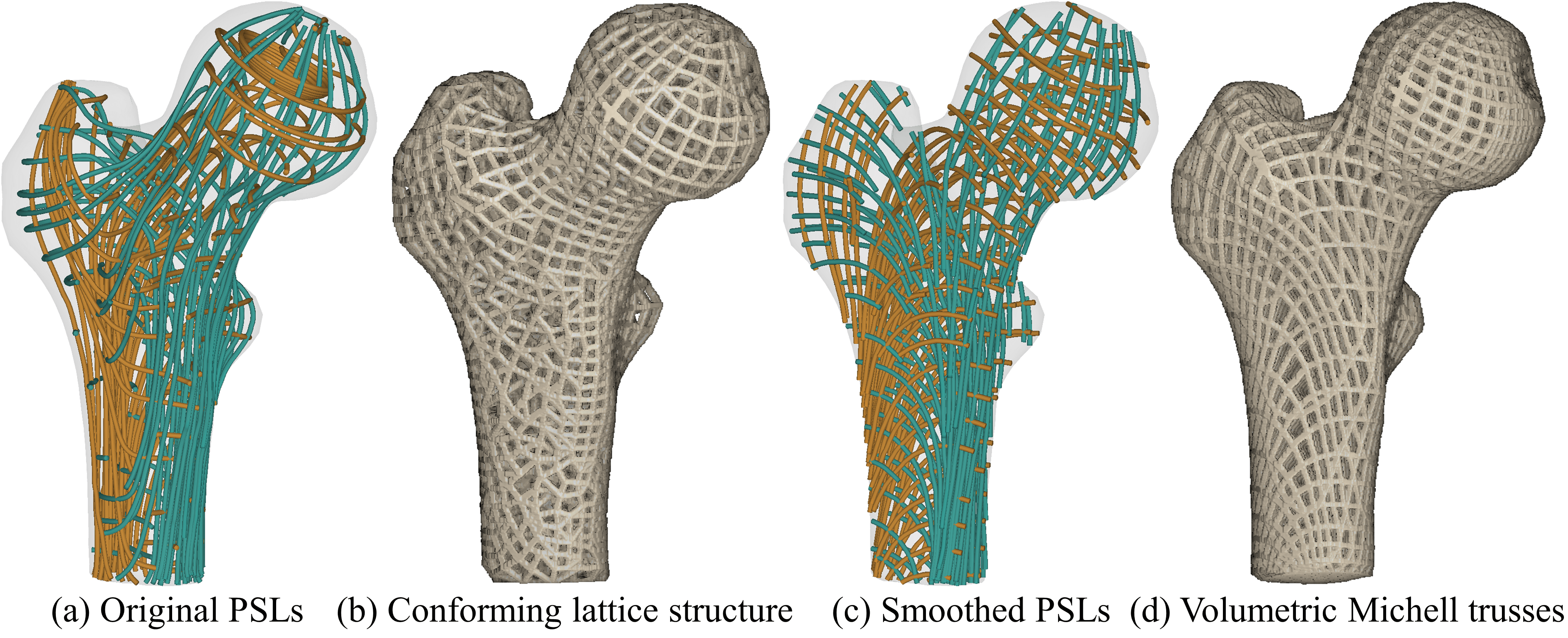}
    \caption{Major (brown) and minor (green) PSLs in the solid object under load (a) and in the smoothed stress field (c), used to generate the designs in (b) and (d), respectively.  
    }
    \label{fig:demo_ConformingLat} 
\end{figure}

SGLDBench directly feds the stress field in the solid domain to the dehomogenization stage, where it is used to generate a stress-aligned lattice structure. Figs.~\ref{fig:demo_ConformingLat}a, b show the initial stress directions in the solid domain and the conforming lattice structure where the cell sizes have been further adapted to the local von Mises stresses. 

The conforming lattice structure addresses the intersection issue found in PSL-guided infills by relaxing the requirement to strictly follow the principal stress directions. This approach builds upon the field-aligned hex-dominant meshing method by Gao \emph{et al.}~\cite{gao2017robust}, which employs an orthogonal frame field to align the edges of a hexahedral mesh with this field. In the conforming lattice structure, the frame field is replaced with the field of principal stress directions, and mechanical anisotropy is incorporated into the field-aligned parameterization process.

Due to the presence of degenerate points and regions in a stress field, it is generally impossible to compute a conforming hexahedral mesh for all but the simplest fields. The conforming lattice structure addresses this limitation by employing a local smoothing strategy. The key idea is to leverage the rotational symmetry of principal stress directions to generate a smoothed direction field, which enables the construction of a conforming hexahedral mesh.

From this smoothed direction field, a position field is computed, ensuring that its gradient aligns with the adjusted stress directions. By combining the smoothed direction field and the position field, the method constructs a graph structure and subsequently extracts the final mesh. To preserve the divergence and convergence properties of the underlying stress field, irregular vertex connections are introduced in the final structure, resulting in an edge-graph representation. 

\subsubsection{Volumetric Michell trusses}
Volumetric Michell Trusses compute a stress-aligned hexahedral lattice using a parametric approach to align truss elements with the principal directions of the stress field.
This method involves two key steps:

\noindent \textbf{Frame Field Smoothing.}
	The algorithm first applies an FEA to compute the stress tensor field, followed by a frame field generation aligned with the principal stress directions.
    However, to achieve a global parametric structure, this step smooths out tensor field singularities, sacrificing local alignment near degenerate points for overall global consistency.
    To address this, the method uses Loubignac iterations~\cite{loubignac1977continuous} to smooth out the discontinuous stress field. This iterative method adjusts the stress field to ensure that it becomes continuous across element boundaries, thereby allowing for smoother and more uniform alignment in subsequent steps.
    Therefore, a smoothness energy function that penalizes sharp changes in frame directions is minimized.
    This optimization enables a globally smooth frame field that approximates the original stress directions.

\noindent \textbf{Tracing Integer Isolines.}
    After smoothing, integer isolines of the volumetric texture parameterization are traced.
	This step maps the truss nodes to integer points of the parameterization, yielding the geometry for the extracted graph structure.
	Its connectivity follows from the nodes' adjacency in the grid.
    To ensure flexibility, the method allows scaling of the parameterization via a user-defined resolution parameter $\rho$, which controls the density of the truss structure.

Due to the applied smoothing of the initial stress field, volumetric Michell trusses produce a regular hexahedral lattice structure with improved continuity of load transmissions, as demonstrated in 
Fig.~\ref{fig:demo_ConformingLat}c, d. However, the smoothing process can substantially alter the initial stress field, leading to significant deviations in the resulting design's stiffness from the optimal value.

\subsection{Infill Voxelization}
\label{subsec:infill}
While TO and porous infill compute a binary material field on a 3D voxel grid, the other methods compute a 3D lattice structure composed of edges and nodes. To enable a meaningful comparison of the structural properties of all approaches, SGLDBench voxelizes these structures into a voxel grid. The grid resolution is selected automatically to represent edges with a minimum required voxels. SGLDBench computes for each edge the intersected voxels via the DDA line drawing algorithm~\cite{amanatides1987fast}. These voxels are set to solid. Edges are thickened by setting for all these voxels their 26 neighboring voxels to solid. 

For all edge-based infill strategies, the material budget and the targeted edge thickness are prescribed.  
The methods are then conducted in a dichotomy manner to find the settings that match closely the design specifications, i.e., the design process is run multiple times to find the design that matches the material consumption under the edge thickness constraint. 

\subsection{Visualization and Layout Analysis} 
Once a Voronoi infill, PSL-guided infill, conforming lattice structures, or volumetric Michell trusses has been computed, users can view the meshes and graph structures using MATLAB's mesh viewing operations, rendering edges as cylinders with a specified width. Thus, voxelizing the infill into a 3D material field is not required.

To compute the compliance of a design, lattice infill designs must first be voxelized into a 3D material field. SGLDBench then uses its linear elasticity solver to perform an FEA and simulate the stress field, from which the compliance is computed. For density-based TO and porous infill optimization, the compliance history throughout the optimization process is recorded and can be visualized via a curve plot, as demonstrated in Fig.~\ref{fig:demo_topopti_GL}.

\begin{figure}[t]
    \centering
    \includegraphics[width=0.98\linewidth, trim=0.0cm 0.0cm 0.0cm 0.0cm, clip=true]{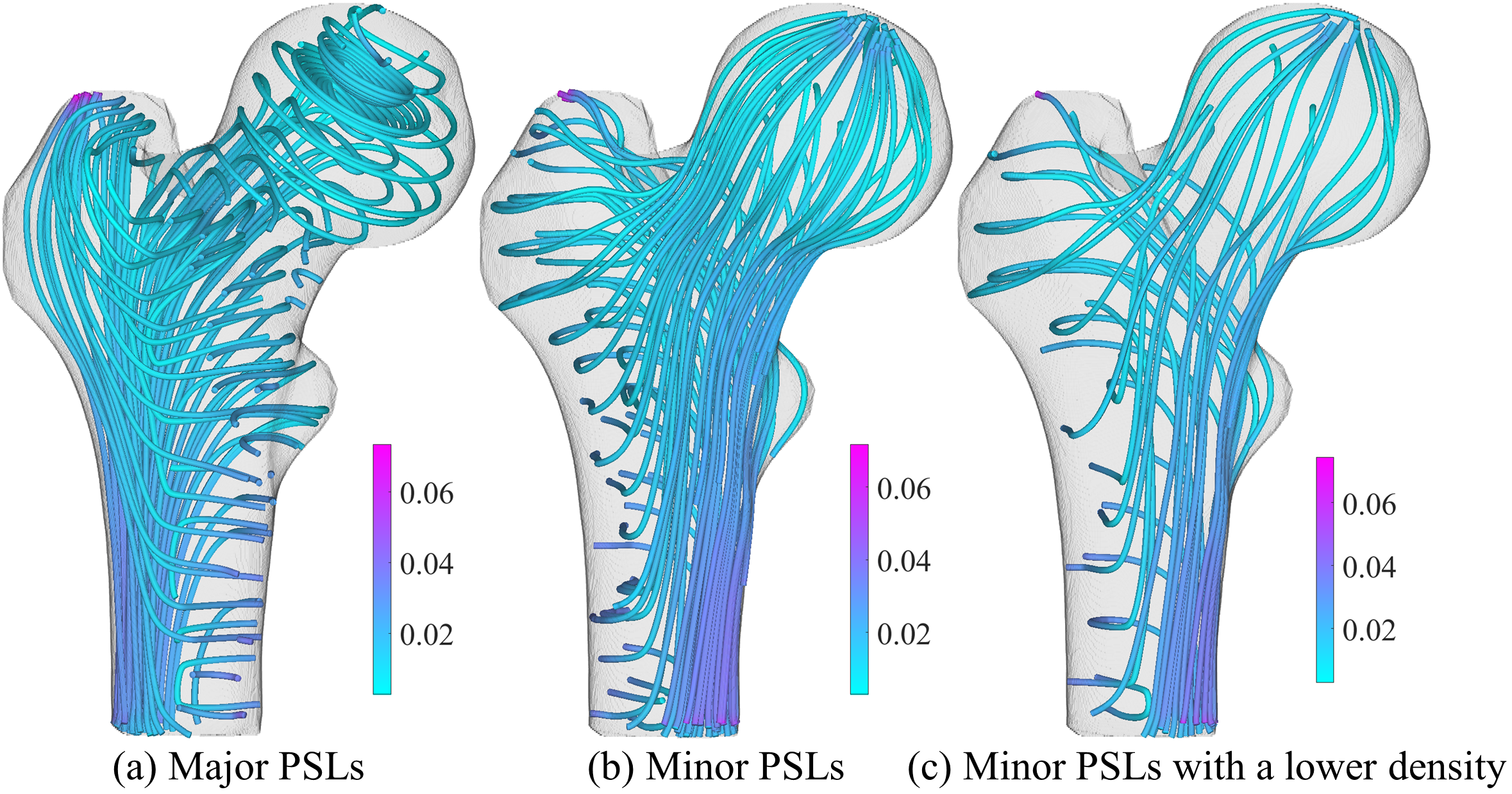}
    \caption{PSL visualization with cylindrical elements and color-coded von Mises stresses using MATLAB visualization programs.}
    \label{fig:PSLsVis}
\end{figure}

To visualize a stress field, an evenly spaced set of PSLs covering the domain as uniformly as possible is computed using MATLAB programs. Users can control the density of seeded PSLs and select scalar stress measures, such as the von Mises norm, to map onto the lines' colors. PSLs can be computed for the initial stress field in the solid domain. Examples of PSL-guided visualizations using MATLAB are shown in Fig.~\ref{fig:PSLsVis}.

For realtime visualization of even high resolution designs, SGLDBench provides an advanced WebGL-based volume visualization module. It performs isosurface and direct volume rendering, and assists users in a stress-based comparative design analysis. Isosurfaces in a 3D material field are rendered using GPU ray-casting, with screen-space ambient occlusion to enhance depth perception, as demonstrated in Fig.~\ref{fig:teaser}. A view-space parallel clip plane can be moved back and forth to expose otherwise occluded structures.

Direct volume rendering is particularly used to visualize the scalar von Mises stress field, enabling the evaluation of whether a material will permanently deform under the given stress state. High magnitudes of the von Mises stress indicate a risk of fracture under the applied loads, and direct volume rendering effectively highlights the spatial regions where this danger is significant. This provides a powerful tool for structural and mechanical analysis. SGLDBench's WebGL interface allows users to color a ray-traced infill surface based on the von Mises stress.

In addition, SGLDBench's visualization module highlights the differences between the principal stress directions in the solid design and the computed infill design. SGLDBench computes the stress field of the infill using the initial boundary conditions and generates an auxiliary grid where each voxel stores a single deviation measure. This measure indicates the deviation of the stress directions corresponding to the principal stresses with the maximum absolute value, based on the ordering of the absolute values of the principal stresses in the initial solid and the infill.

The resulting scalar field is visualized through direct volume rendering. SGLDBench employs a transfer function that maps directional deviations linearly to colors, ranging from white (low deviation) to light brown (medium deviation) and dark brown (high deviation). Opacity is initially set to one but can be adjusted by the user to smoothly fade out regions with low or high deviation.

Given an optimized infill structure, SGLDBench can also be used to apply new boundary conditions to it, allowing users to probe conditions different from those for which the structure was initially optimized. This functionality is inspired by worst-case structural analysis~\cite{zhou2013worst, zhang2022large}, a method used to evaluate the performance and reliability of a structure under its most unfavorable conditions. While SGLDBench is designed for a completely different use case and cannot perform such analysis, it provides the tools to explore similar scenarios.

Specifically, SGLDBench includes an interface to modify the initial boundary conditions by changing the direction of forces and re-computing the compliance and von Mises stress under the new conditions. A visualization of the structure with stress-based color coding highlights the mechanical strengths and weaknesses of different parts, enabling a deeper understanding of its behavior under varied conditions.

\section{Experiments} \label{sec:experiments}
We demonstrate the use of SGLDBench to generate and analyze lightweight designs for various models and boundary conditions. The models include a human femur (\emph{Bone}), a machine part commonly seen in engineering applications (\emph{Part}), and \emph{Cantilever}, a widely used benchmark model in TO. All models are initially provided as triangle meshes. We showcase the usability of SGLDBench with additional datasets in the supplementary material.

All experiments are conducted on a desktop computer equipped with an Intel 6-core Xeon W2235 CPU, 64 GB of RAM, and an NVIDIA RTX 2070 GPU with 8 GB of video memory. We intentionally select a mid-range architecture to demonstrate SGLDBench's capabilities on affordable hardware. The main memory limits the maximum number of simulation elements to approximately 160 million for simulating a 3D stress field. 

\subsection{Performance Evaluation}   
\noindent \textbf{Solver. } For the \emph{Cantilever} model at a voxel grid resolution of 860×430×430 (159 million elements), SGLDBench solves the FEA linear system on the CPU in roughly 30 minutes, achieving convergence within 41 solver iterations at a threshold of $1.0\times10^{-3}$

A one-to-one CUDA implementation of the solver by Wu~\emph{et al.}~\cite{wu2015system} on the RTX 2070 GPU can simulate up to 45 million elements before running out of GPU memory. For this number of elements, the GPU implementation needs 91 seconds to solve the FEA linear system. SGLDBench requires 610 seconds for the same setting, showing roughly a 7x reduction of the performance compared to the optimized GPU solver. For the used GPU this is a reasonable reduction, and slightly better than commonly reported when GPU and CPU implementations of similar problems are compared. 

\begin{figure}[t]
    \centering
    \includegraphics[width=0.98\linewidth, trim=0.0cm 0.0cm 0.0cm 0.0cm, clip=true]{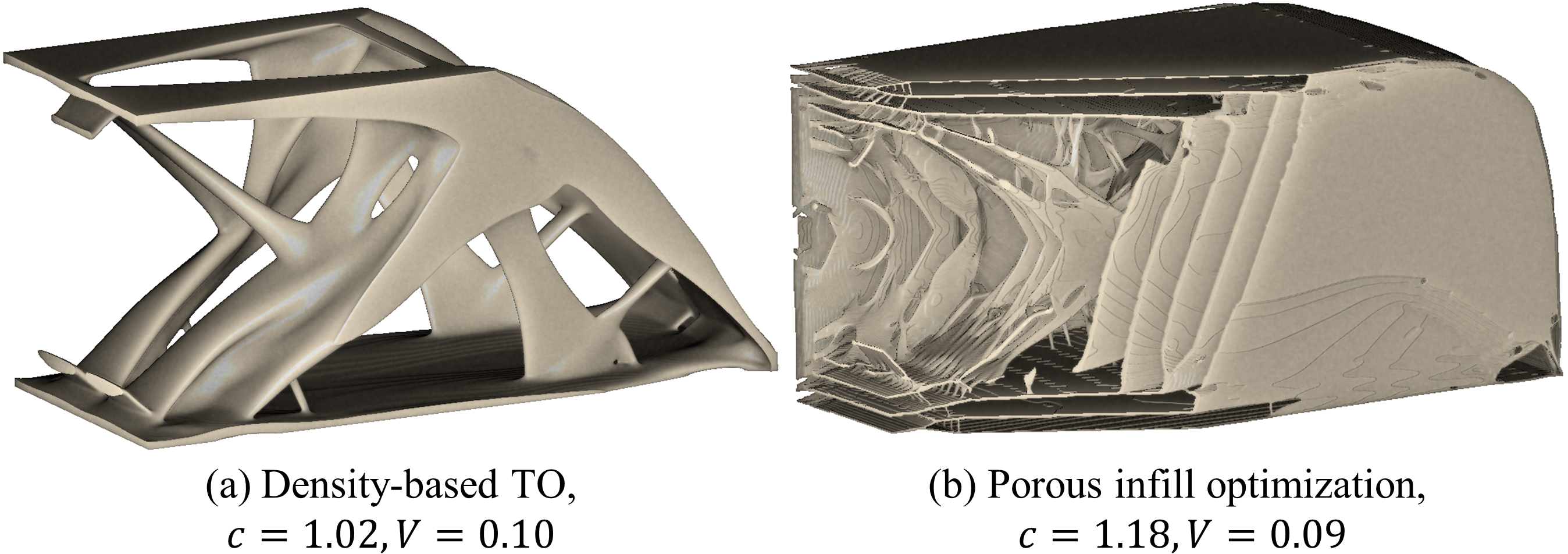}
    \caption{Material fields generated with TO (a) and porous infill optimization (b) for \emph{Cantilever} using a $800\times400\times400$ voxel grid. 
    The left face of the cubic design domain is fixed, and a downward force acts along the bottom-right edge.}
    \label{fig:highres}
\end{figure}

\noindent \textbf{Iterative Optimization. } TO and porous infill optimization require additional memory for updating the material distribution during numerical optimization. SGLDBench performs these optimizations with grids of up to 130 million simulation elements, corresponding to a 800×400×400 simulation grid with 386 million degrees of freedom. For these cases, SGLDBench completes each optimization iteration in approximately 45 minutes for TO and 67 minutes for porous infill optimization. The results shown in Fig.~\ref{fig:highres} are generated using 30 iterations of TO and 280 iterations of porous infill optimization.

Compared to an optimized OpenMP CPU implementation of TO~\cite{traff2023simple}, SGLDBench is only about 1.6 times slower when reproducing the same \emph{Cantilever} model at a resolution of 640×320×320 on a similar computing architecture (a 48-core Xeon CPU). This demonstrates the efficiency of the MATLAB computing environment in combination with SGLDBench's element-wise computation structure.

\begin{figure*}[t]
    \centering
    \includegraphics[width=0.98\linewidth, trim=0.0cm 0.0cm 0.0cm 0.0cm, clip=true]{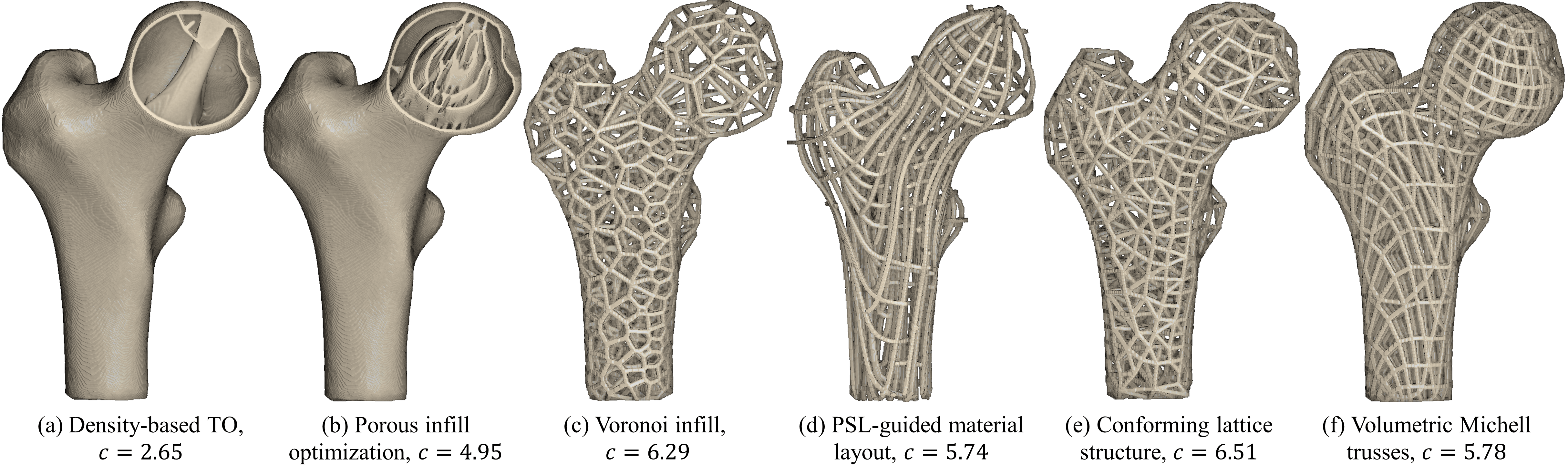}
    \caption{Same designs as in Fig.~\ref{fig:teaser} but with lower material budget of roughly 0.22. 
    }
    \label{fig:lowvolume}
\end{figure*}

All designs for \emph{Bone} in Fig.~\ref{fig:teaser} are generated with a $384\times256\times512$ voxel grid, corresponding to 30 million degrees of freedom. TO and porous infill optimization, respectively, generate the infill in 1.5 hours and 15.1 hours. Porous infill requires a significantly higher number of optimization iterations and additional computations to enforce the local volume constraint. Due to the high geometric complexity of porous infills, generating these infill also requires a higher number of iterations for solving the linear FEA system in each optimization iteration.  

\noindent \textbf{Lattice Infill Optimization. } 
To compute the various types of lattice infills, we utilize state-of-the-art implementations currently available, which are provided as either C/C++ codes (PSL-guided layouts, Voronoi, and conforming lattice infills) or MATLAB programs (Michell trusses). Executing these implementations through SGLDBench does not result in performance penalties.

SGLDBench requires approximately 10 minutes to generate the Voronoi infill and the PSL-guided infill, with about half of this time spent simulating the 3D stress field in the initial solid domain. Using this stress field, the external codes for generating the conforming lattice infill take around 5 minutes, while the MATLAB code for generating the volumetric Michell truss requires approximately 4.5 hours.
For the latter, slightly less than half of the execution time is allocated to smoothing the original stress field.

\subsection{Material Use as a Modelling Parameter}
To evaluate the response of different design methods to changes in material consumption, we repeat all experiments shown in Fig.~\ref{fig:teaser} with a reduced volume fraction. Instead of the original volume fraction of 0.4, SGLDBench now uses a lower volume fraction of approximately 0.2, enforcing finer and less dense support structures. For methods that generate graph structures, the thickness of the voxelized edges remains constant.

The results, shown in Fig.~\ref{fig:lowvolume}, reveal an increased sparseness in all designs, accompanied by reduced stiffness and significantly varied topologies. TO and porous infill methods require no changes to their simulation parameters but must rerun the entire optimization process to produce the results. In contrast, other approaches can utilize the stress field from the initial solid domain and need to rerun only the steps that generate the graph structure from it.

With a lower material budget, the PSL-guided infill demonstrates surprisingly good relative performance, as it effectively utilizes the material to generate support structures along the most dominant stress directions. 

One possible reason for this reversal is that the resolution of the conforming lattice is too low, leading to the misalignment of many edges in the graph structure with the dominant stress directions. Furthermore, adaptive porosity also plays a critical role in reducing compliance. Specifically, in regions of high stress, more material should be allocated to resist strain effectively. This could explain the relatively moderate stiffness of the Voronoi infill, which is unable to concentrate more material in high-stress regions due to the low material budget. Moreover, by design, the Voronoi infill does not align with the dominant stress directions, further contributing to its reduced mechanical performance.

The stress field in \emph{Bone} is relatively simple and contains few degeneracies, allowing the Michell trusses to preserve most features effectively. This behavior, however, changes with \emph{Part} (see the first row of Fig.~\ref{fig:Part1}). In this case, the smoothed stress field exhibits significant differences from the initial field, causing many infill edges to deviate substantially from the original stress directions. As a result, the Michell trusses demonstrate lower stiffness compared to the other alternatives.

\subsection{Infill Resolution and Compliance}
To explore the relationship between compliance, volume fraction, and geometric infill resolution, SGLDBench is used to evaluate the stiffness of high-resolution lattice infills for \emph{Bone}. The volume fraction, edge thickness, and voxel grid resolution are user-defined parameters, and SGLDBench computes the compliance of the infill using a single FEA iteration.

\begin{figure}[ht]
    \centering
    \includegraphics[width=0.9\linewidth, trim=0.0cm 0.0cm 0.0cm 0.0cm, clip=true]{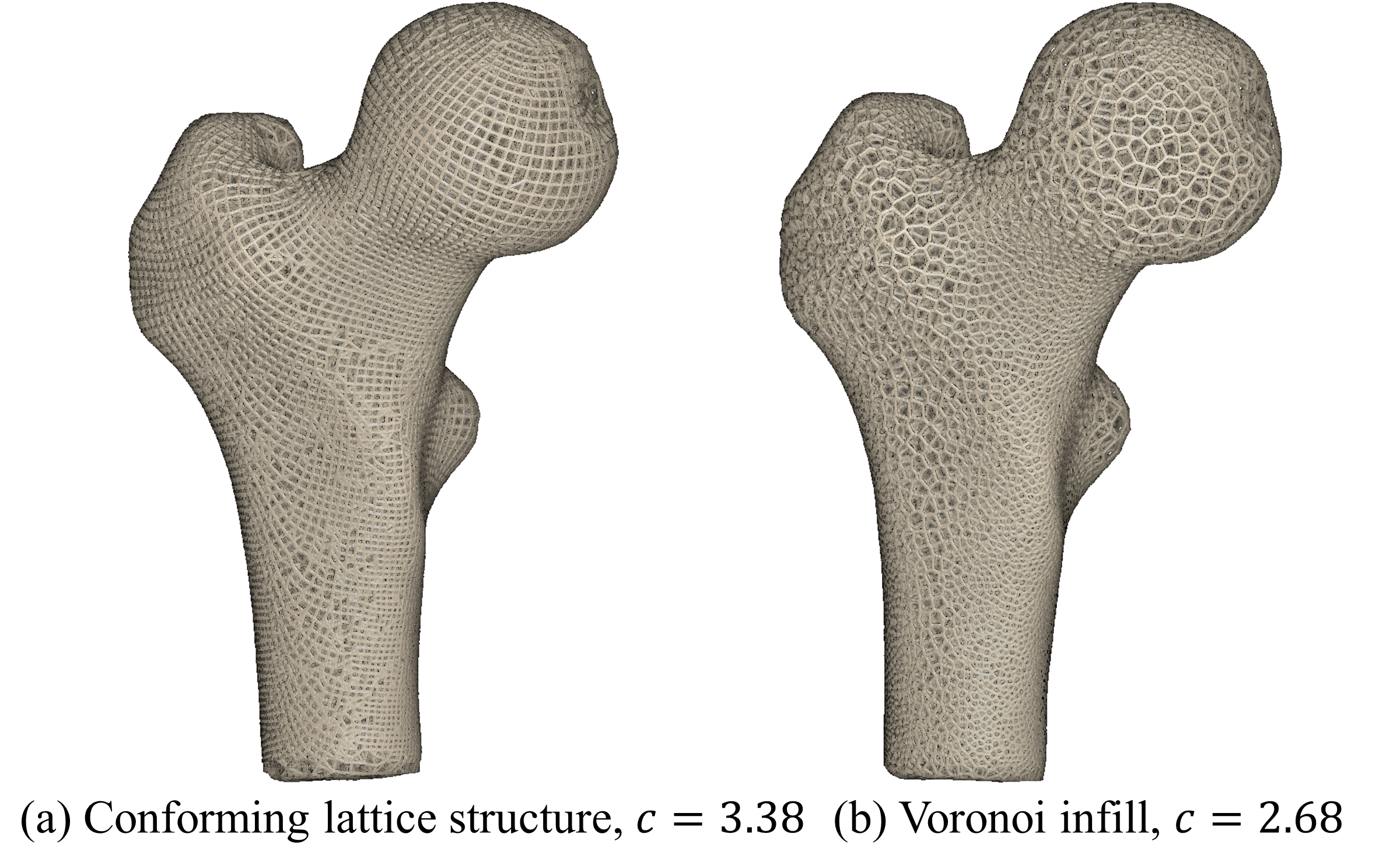}    
    \caption{High resolution results using the same boundary conditions and material budget as in Fig.~\ref{fig:teaser}.
    }
    \label{fig:highresLattice}
\end{figure}

In Fig.~\ref{fig:highresLattice}, a voxelized conforming lattice infill and Voronoi infill with $136,877$ and $396,458$ edges, respectively, are shown, using the same boundary conditions as in Figs ~\ref{fig:teaser} and ~\ref{fig:lowvolume}.
SGLDBench generates a 848×576×1200 voxel grid and constructs a finite element model from it. The voxelized versions of the infills are visualized using SGLDBench's WebGL visualization module, employing isosurface ray-casting and ambient occlusion for enhanced depth perception.

To compute the compliance of the voxel models, approximately 380 million degrees of freedom are solved, which SGLDBench completes in roughly 45 minutes. Interestingly, as shown in Fig.~\ref{fig:teaser}, the stiffness appears to be independent of the geometric details of the infills when using the same material budget.

\subsection{Infill Geometry}
\begin{figure}[t]
     \centering
     \includegraphics[width=1.0\linewidth, trim=0.0cm 0.0cm 0.0cm 0.0cm, clip=true]{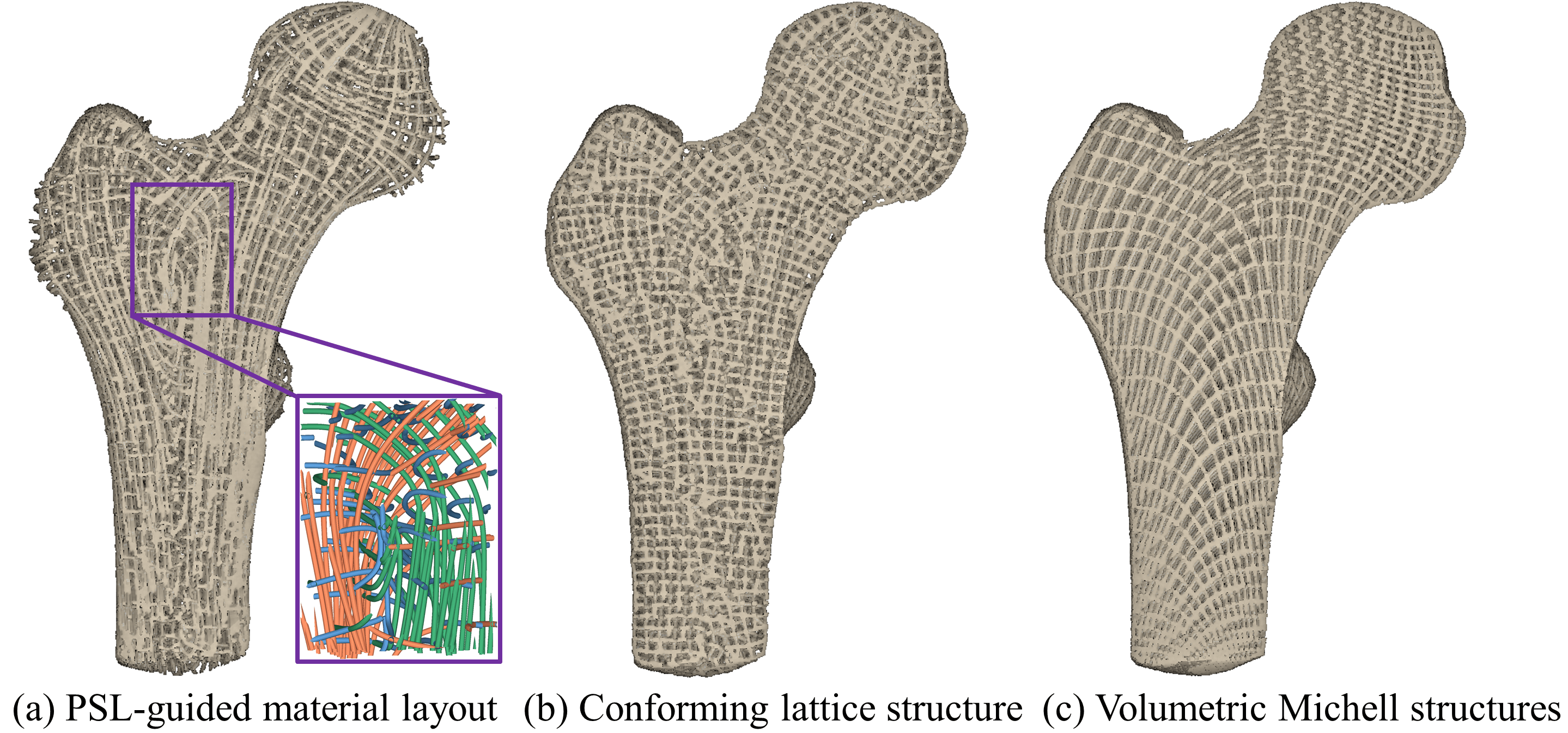}
     \caption{A clip plane reveals the interior structure of different designs. The closeup view shows the PSLs in the selected region, with the major, medium, and minor PSLs, respectively, shown in ocher, green, and blue.}
     \label{fig:clipplane}
\end{figure}

Given the varying compliance of different types of infills for the same object and load conditions, we use SGLDBench to investigate the major causes of these differences. We compare the interior structures of graph-based infillS, as illustrated with \emph{Bone} in Fig.~\ref{fig:clipplane}. A clip plane is used to expose the interior regions of the PSL-guided infill, the conforming lattice structure, and the volumetric Michell trusses.

The PSL-guided infill serves as a reference, illustrating the primary load transmission pathways. Overall, the edges of the conforming lattice structure align with the PSLs. However, in regions where the curvature of the PSLs changes significantly (see the inset in Fig.~\ref{fig:clipplane}), the lattice loses its geometric regularity. These irregularities appear to be associated with degeneracies in the stress field, where the principal stress directions flip.

In contrast, the volumetric Michell trusses exhibit a highly regular geometric structure but demonstrate less alignment with the original stress directions. This discrepancy arises because the original stress field is smoothed, which alters the stress field globally and particularly eliminates unwanted degeneracies.

The visualization of the 3D conforming lattice structure reveals several stitching edges in the interior that do not align with any of the dominant stress directions. Additionally, the number of longer edge sequences that consistently follow one of the PSLs is lower compared to the PSL-guided infill and the volumetric Michell trusses.
This behavior might be attributed to the approach allocating too much material to maintain a consistent edge graph rather than prioritizing stiffness. Oscillating load paths, represented by inconsistent edge sequences, may lead to less effective load transitions and, consequently, reduced stiffness.

\subsection{Variable Load Structural Analysis}
To shed light on how well a design optimized for a specific load case can resist forces applied from a different direction, SGLDBench is used in the following way: First, all six infill design methods are applied with the same boundary condition to compute six different designs, and the von Mises stress field in the optimized material field is computed. Then, the applied forces are changed, and the von Mises stress field is recomputed for all designs. The designs, color-coded with the von Mises stress under the varied forces, are shown in Fig.~\ref{fig:Part1}.

Density-based TO consistently shows higher compliance under the new force conditions. While the compliance increases only slightly with a slight change of the force direction, larger directional changes result in a significant loss in stiffness. The coloring with the von Mises stress shows where the critical structures occur under the new forces and have the potential to break. A significant re-distribution of the von Mises stresses indicates a significant loss in stiffness. 

All other designs show the same slight change in stiffness as porous infill when the force direction changes only slightly. For more significant changes in the force directions, however, all other designs perform significantly better than the one optimized with density-based TO. Interestingly, even the PSL-guided material layout, which is specifically aligned with the major stress directions occurring with the initial force setting, can significantly better resist the new force directions. 

Designs that show a more uniform distribution of material throughout the domain, such as the Voronoi infill, the conforming lattice structures, and the volumetric Michell trusses, are much better at resisting varying forces compared to topology-optimized designs. When the force directions are varied significantly from their initial values, in some cases the stiffness becomes even higher. These are situations where the new forces are along some of the principal stress directions that have occurred in the initial stress field, and along which some of the structures have grown, so that the structure is bearing more normal stresses than shear stresses. 

\subsection{Reproducing Stresses in the Solid Domain}
\begin{figure*}[t]
    \centering
    \includegraphics[width=0.95\linewidth, trim=0.0cm 0.0cm 0.0cm 0.0cm, clip=true]{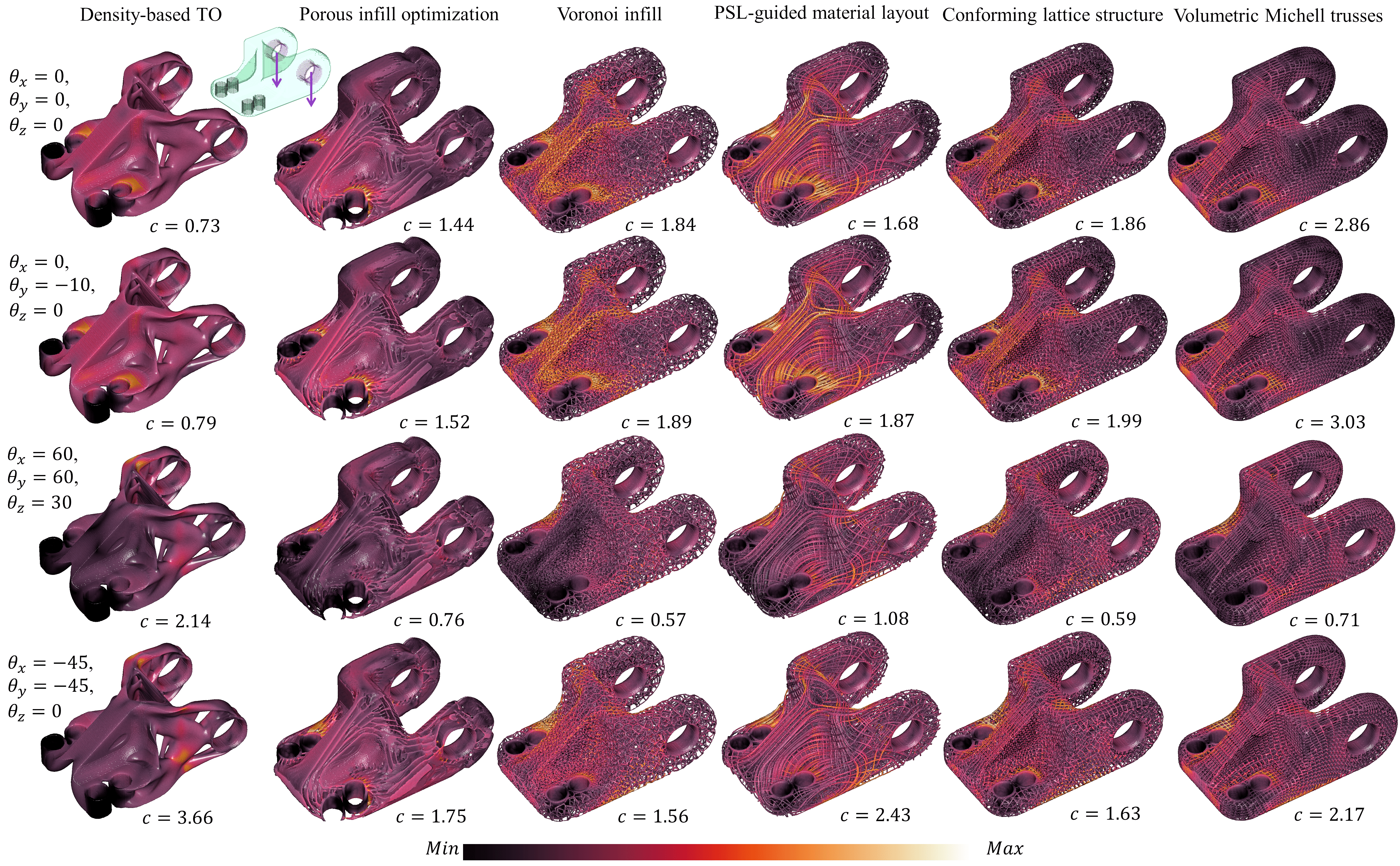}
    \caption{Variable load structural analysis. Top row: All infills are optimized for the same boundary conditions (see inset with downward forces). Rows 2–4: Varying forces (Euler angles $\theta_x, \theta_y, \theta_z$ indicate angular deviation from the downward direction) are applied to the optimized infills. All designs are shown with their compliance $c$ and are color-coded according to the von Mises stress.}
    \label{fig:Part1}
\end{figure*}

Using \emph{Part} shown in Fig.~\ref{fig:Part1}, SGLDBench is applied to analyze how well different designs reproduce the stress directions in the initial solid configuration. This analysis allows users to examine, for instance, whether force transmission through tension and compression occurs along the same load paths as in the initial solid. Since an infill design can, in principle, transmit forces along the most efficient load paths in the solid design or diverge to alternative paths, understanding the relationship between compliance and the reproduction of stress directions can provide valuable insights for improving infill designs.

\begin{figure*}[t]
    \centering
    \includegraphics[width=0.98\linewidth, trim=0.0cm 0.0cm 0.0cm 0.0cm, clip=true]{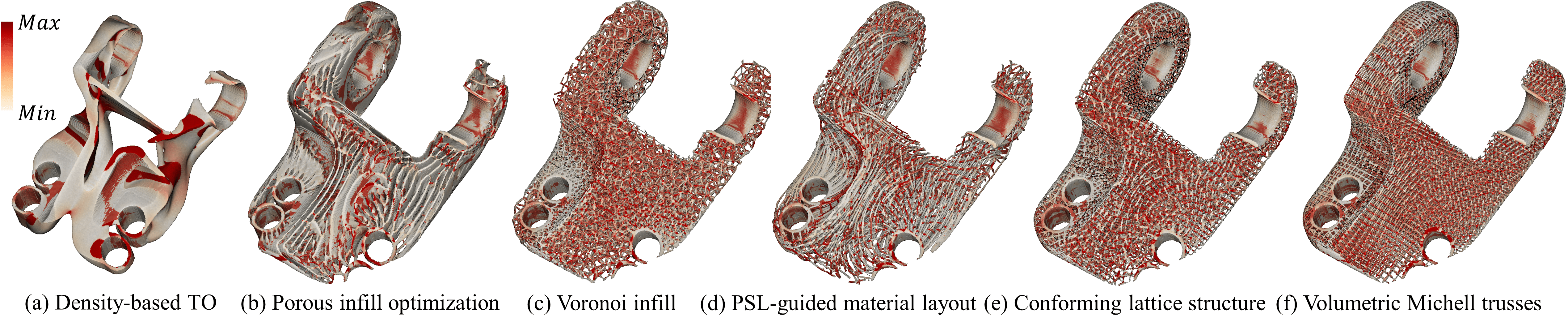}
    \caption{Stress-to-stress alignment. High (white) to low (brown) alignment between stresses in the solid and the optimized infill with respect to the same boundary conditions is shown. A view plane aligned clip plane reveals interior parts.}
    \label{fig:stresscompare-1}
\end{figure*}

Fig.~\ref{fig:stresscompare-1} shows visualizations of the alignment fields computed for each infill. Iso-surface rendering with the color transfer function described in Sec.~\ref{subsec:infill} is used to emphasize regions with high deviation in brown. 
TO produces an infill that, in many regions, aligns well with the major stress directions in the solid under load. However, significant deviations are observed in certain areas, particularly in thin structures and regions near the fixed elements where stress concentrations are highest.

Similarly, porous infill exhibits more pronounced deviations, which are distributed throughout the entire domain. This behavior arises from its space-filling material distribution, which emphasizes uniformity rather than alignment with the stress directions.

The stress deviations are highest in the Voronoi infill, as the initial stress directions are not considered in its edge-graph layout. Only when edges align with boundary elements, which are incorporated into the stress analysis of both the solid object and the infill, stress deviations are lower.

The material in the PSL-guided infill is distributed along the principal stress directions, resulting in good alignment with the initial stress field. However, as shown in Fig.~\ref{fig:stresscompare-1}, alignment accuracy decreases at the boundaries of the voxelized edges. This reduction in accuracy stems from discretization-induced inaccuracies in the stress simulation—a limitation shared by all graph-based approaches when voxelized structures are used for compliance analysis.

Additionally, all graph-based infills contain numerous edges aligned with the two principal stress directions not corresponding to the direction of the maximum absolute stress value. This can lead to high directional deviations along these edges.

The deviation maps of the conforming lattice structure and the Michell trusses are very similar, showing the same low deviation patterns in the boundary region as the Voronoi infill. In both infills, there is still a considerable number of voxels with significantly different dominant stress directions than the solid object.

\section{Conclusion and Outlook}
SGLDBench is a benchmark suite designed for the simulation and analysis of stiff, lightweight designs, with a special emphasis on high-resolution 3D models. It supports the computation of six distinct design types, employing methods that range from purely stiffness-based optimization to geometry-aware infill generation.

SGLDBench enables users to create designs at varying resolutions and material consumption levels while assessing their mechanical and geometric properties. Various visualization options provide additional insights into governing stress states and a design's geometric structure. The resistance of a design to new force situations can be assessed via the color coding of designs with scalar stress measures. Additionally, SGLDBench visualizes deviations between stress directions in the initial solid object and the generated design, providing insights into relationships between stiffness, geometric properties, and stress replication.

The suite allows users to compute individual designs with specific boundary conditions and offers flexibility for integrating new design strategies into its publicly available code base. Moreover, SGLDBench supports the creation of novel design types by combining existing strategies. For instance, the material field of an optimized design can be computed via TO and downloaded to extract its boundary as an isosurface. This surface can then be uploaded to SGLDBench to compute an infill restricted to the interior of the surface, as demonstrated in the supplemental material.

By leveraging SGLDBench's capabilities, several intriguing properties of existing design strategies have been uncovered, paving the way for new research directions in lightweight design. For example, 
why does the mechanical performance of conforming lattice structures fall below expectations in 3D domains, what role does adaptive porosity play in achieving high stiffness, what is the interplay between truss thickness, cell size, and stiffness in lattice infills, how can a tensor field be optimally smoothed to minimize stiffness deviation while maintaining efficient load transmission paths, or how would a Voronoi infill perform if additional constraints on edge stress reproduction were applied?
SGLDBench provides researchers with tools to investigate these questions and develop improved design strategies tailored to diverse objectives.

\section*{Acknowledgments}
This work was supported by the German Research Foundation (DFG) under grant number WE 2754/10-1. Several open-source projects are used when developing SGLDBench, including the volumetric Michell trusses~\cite{arora2019volumetric}, conforming lattice structure~\cite{Wu2021TVCG}, tetrahedral mesh generators \emph{TetGen}~\cite{hang2015tetgen} and \emph{fTetWild}~\cite{hu2020fast}, voxelization functions~\cite{Adam2024}, and MMA implementations~\cite{svanberg1987method, aage2013parallel}. Hereby, we sincerely acknowledge the generosity of the authors in making their code publicly available.

\ifCLASSOPTIONcaptionsoff
  \newpage
\fi

\bibliographystyle{IEEEtran}
\bibliography{IEEEabrv,bibs}


\vspace{-30pt}

\begin{IEEEbiography}
    [{\includegraphics[width=1in,height=1.25in,clip,keepaspectratio]{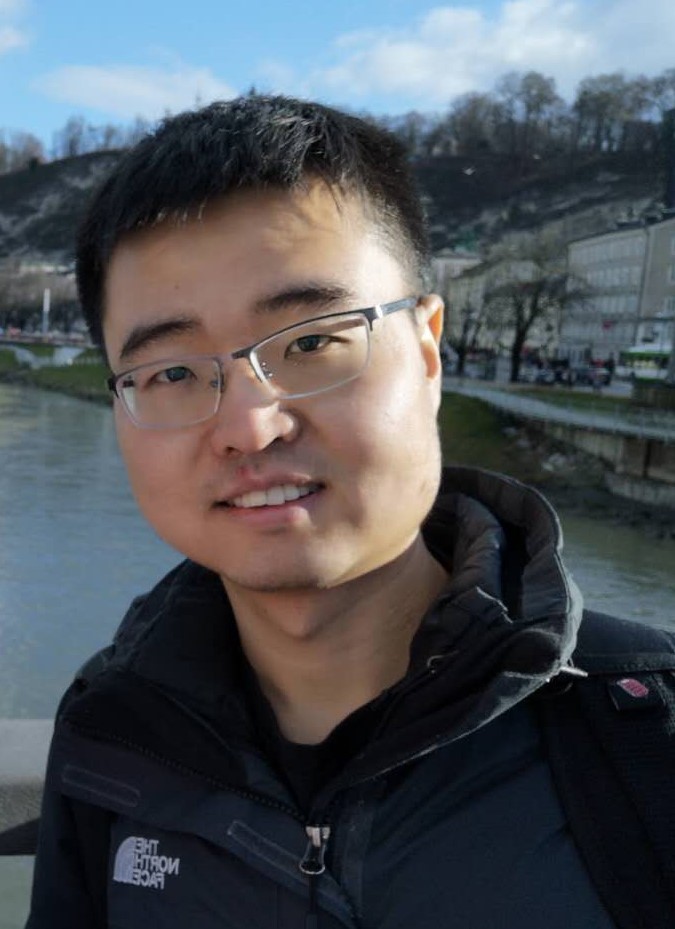}}]{Junpeng Wang}
    is a Post-doc in the Computer Graphics and Visualization Group at Technical University of Munich, Germany. He obtained his PhD from the same group in 2023. His research interests range from scientific visualization to computational design, with a special focus on lightweight structure design and optimization.
\end{IEEEbiography}

\vspace{-30pt}

\begin{IEEEbiography}
    [{\includegraphics[width=1in,height=1.25in,clip,keepaspectratio]{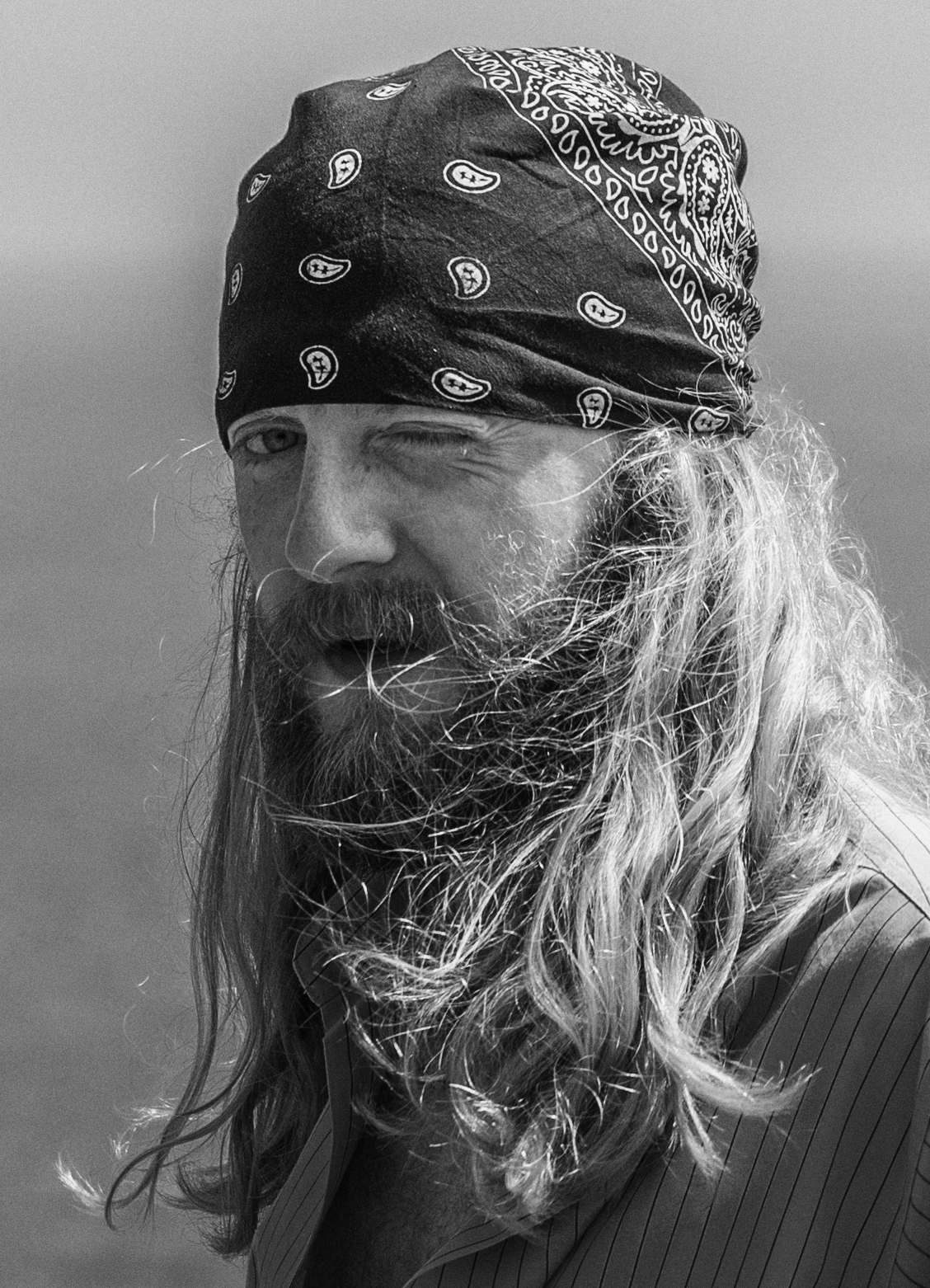}}]{Dennis R. Bukenberger}
    is a postdoc in the Computer Graphics and Visualization Group at the Technical University of Munich, Germany.
    Before this, he was a postdoc at the Technical University of Dortmund.
    He obtained his PhD at the University of Tübingen in 2021.
    His research interests include meshing algorithms, Voronoi diagrams, geometry processing \& sustainable design engineering.
\end{IEEEbiography}

\vspace{-30pt}

\begin{IEEEbiography}
    [{\includegraphics[width=1in,height=1.25in,clip,keepaspectratio]{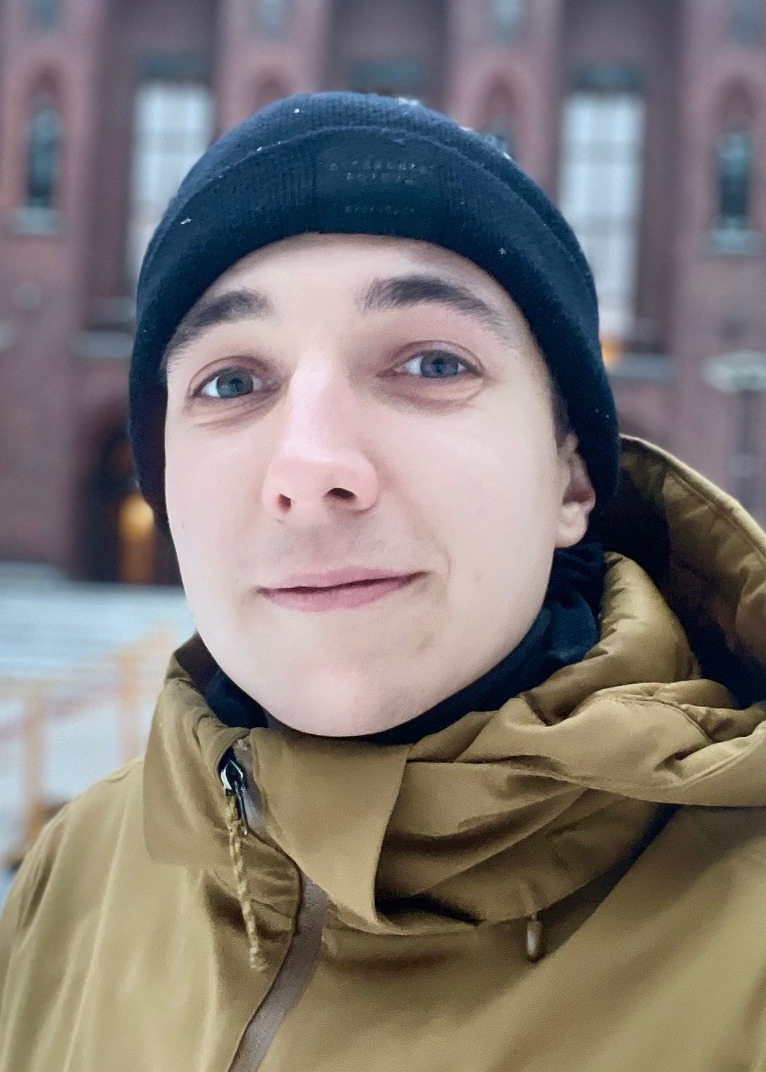}}]{Simon Niedermayr}
    is a PhD candidate at the Computer Graphics and Visualization Group at the Technical University of Munich (TUM). He received his Master's degrees in computer science from TUM in 2022. His research interests include differentiable rendering, real-time rendering and  high performance GPGPU programming.
\end{IEEEbiography}

\vspace{-30pt}

\begin{IEEEbiography}
    [{\includegraphics[width=1in,height=1.25in,clip,keepaspectratio]{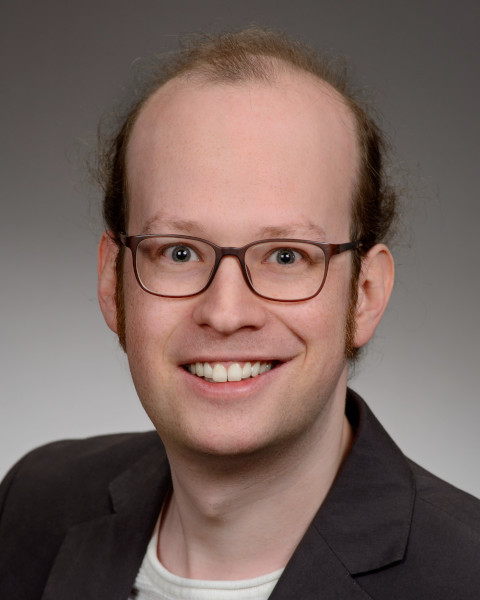}}]{Christoph Neuhauser}
    is a GPU Software Optimization Engineer at Intel. He received his Bachelor's and Master's degrees in computer science from Technical University of Munich (TUM) in 2019 and 2020. He received his PhD in 2025 at the Computer Graphics and Visualization Group at TUM. Major interests in research comprise real-time rendering, GPU computing and scientific visualization.
\end{IEEEbiography}

\vspace{-30pt}

\begin{IEEEbiography}
    [{\includegraphics[width=1in,height=1.25in,clip,keepaspectratio]{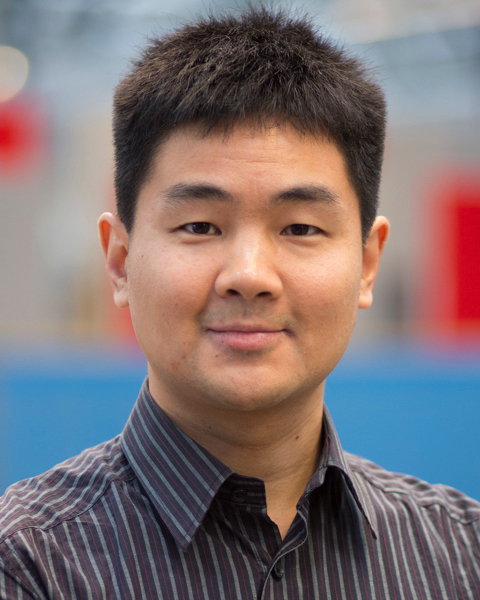}}]{Jun Wu} 
    is an associate professor at the Department of Sustainable Design Engineering, Delft University of Technology. Before this, he was a Marie Curie postdoc fellow at the Department of Mechanical Engineering, Technical University of Denmark. He obtained a Ph.D. in Computer Science in 2015 from TUM and a Ph.D. in Mechanical Engineering in 2012 from Beihang University, Beijing, China. His research is focused on computational design and digital fabrication, with an emphasis on topology optimization.
\end{IEEEbiography}

\vspace{-30pt}

\begin{IEEEbiography}
    [{\includegraphics[width=1in,height=1.25in,clip,keepaspectratio]{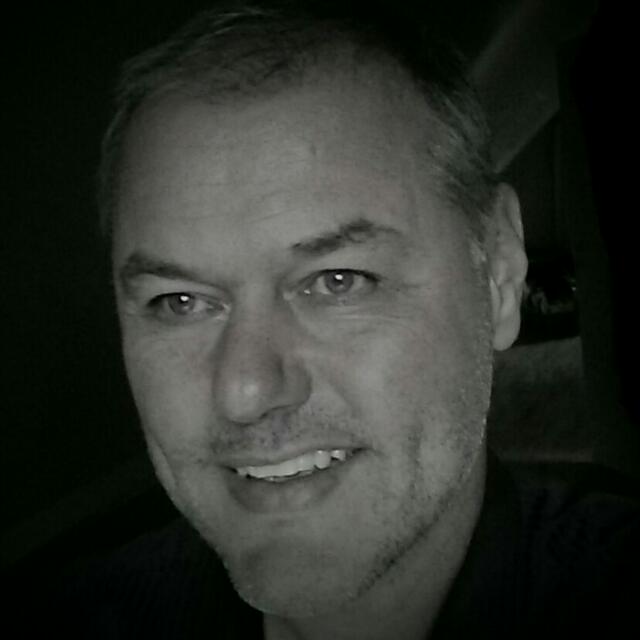}}]{R\"udiger Westermann}
    is a full professor for Computer Science in the TUM School of Computation, Information and Technology. He is the head of Computer Graphics and Visualization group. His research interests include scalable data visualization and simulation algorithms, GPU computing, TO, neural representations for computer graphics and visualization, as well as real-time and photorealistic rendering.
\end{IEEEbiography}
\vfill

\end{document}